\documentclass[12pt,preprint]{aastex}
\usepackage{natbib}
\def\apj{ApJ\,}
\def\apjl{ApJ\,}
\def\aap{A\&A\,}
\def\mnras{MNRAS\,}
\def\aj{AJ}
\def\apjs{ApJS}
\def\cjaa{Chinese J. Astron. Astrophys.}

\def\pre{Phys.~Rev.~E}
\begin{document}
\shorttitle {Luminosity function for galaxies}
\shortauthors{Zaninetti}
   \title{A new luminosity function for galaxies as given 
          by  the mass-luminosity relationship  }
\author{Zaninetti Lorenzo }

\affil {Dipartimento di Fisica Generale, Via Pietro Giuria 1,\\
           10125 Torino, Italy}

\email   {zaninetti@ph.unito.it  \\ 
\url     {http://www.ph.unito.it/$\tilde{~}$zaninett}}

\begin{abstract}
    The search for a luminosity function for galaxies both alternative
    or companion to a  Schechter function is a key problem
    in the reduction of data from catalogs of galaxies.
    Two luminosity functions for galaxies can be built starting from
    two distributions of mass as given by the fragmentation.
    A first overall distribution  function is the Kiang function
    that represents a useful description 
    of  the area and volume distribution 
     of the Poisson Voronoi diagrams.
    The second distribution, that covers the case of low mass 
    galaxies,  is the 
    truncated Pareto distribution  : in this model we 
    have a natural bound due to the   minimum  mass/luminosity
    observed and an upper bound  ( function of the 
    considered environment)   represented 
    by the boundary with the observed mass/luminosity overall
    behaviour. 
    The mass distribution is then converted into a
    luminosity  distribution through  a standard mass-luminosity
    relationship.
    The mathematical rules to convert the probability density
    function are used  and the two
    new functions are  normalised to the total number of galaxies
    per unit volume.
    The test of the two  new luminosity functions for galaxies 
    that cover different ranges in magnitude
    was made on 
    the Sloan Digital Sky
    Survey (SDSS)
    in five different bands;
    the  results are comparable to those of
    the Schechter function.
    A new parameter that indicates
    the stellar content is derived.
    The joint distribution in  red-shift and flux , 
    the mean
    red-shift and the number density connected with 
    the first
    luminosity function for galaxies are obtained by  analogy
    with the Schechter function.
    A new formula
    that allows us to express the mass as a function of the
    absolute magnitude  is derived.
\end {abstract}

\section{Introduction}

Over the years the search for   a luminosity function
for galaxies  has played a relevant role  in
the analysis of data from catalogs.
A model for the luminosity of galaxies
is  the Schechter function
\begin{equation}
\Phi (L) dL  = (\frac {\Phi^*}{L^*}) (\frac {L}{L^*})^{\alpha}
\exp \bigl ( {-  \frac {L}{L^*}} \bigr ) dL \quad  ,
\label{equation_schechter}
\end {equation}
where $\alpha$ sets the slope for low values of $L$ , $L^*$ is the
characteristic luminosity and $\Phi^*$ is the normalisation.
This
function  was suggested  by \citet{schechter} in order to
substitute other analytical expressions, see  for example
formula~(3) in~\citet{kiang1961}.
Over the years this function has also been applied
to describe physical  quantities  related to the optical
luminosity  , such as   the  CO luminosity
for galaxies , \cite{Keres2003},  and the
barionic mass function of galaxies , \cite{Bell2003b}.

An astronomical form of
equation~(\ref{equation_schechter}) can be deduced by introducing the
distribution in absolute magnitude
\begin{eqnarray}
\Phi (M)dM=&&(0.4  ln 10) \Phi^* 10^{0.4(\alpha +1 ) (M^*-M)}\nonumber\\
&& \times \exp \bigl ({- 10^{0.4(M^*-M)}} \bigr)  dM \quad  ,
\label{equation_schechter_M}
\end {eqnarray}
where $M^*$ is the characteristic magnitude as derived from the
data.
This distribution has a maximum at
\begin{equation}
M_{p,max} =
{\it M^*}- 1.085\,\ln  \left( \alpha+ 1.0 \right)
\quad ,
\label{maximummag}
\end {equation}
where ${p,max}$ means position of the maximum.
In approaching this value the function will progressively flatten.

At present this function  is widely  used and
Table~\ref{parameters} reports  , as  an example ,
 the parameters
from  three   catalogs
\begin{itemize}
\item  The 2dF Galaxy
       Redshift Survey (2dFGRS) based on a sample of 75589  galaxies,
       see first line of Table~3 in  \cite{Madgwick_2002}.

\item  The $r^\ast$-band luminosity function for a sample of 147986
       galaxies at $z=0.1$  from the  Sloan Digital Sky
       Survey (SDSS) , see~\cite{Blanton_2003}.
\item  The  galaxy luminosity function
        for a sample of  10095 galaxies
        from the Millennium Galaxy Catalogue (MGC)
        , see~\cite{Driver2005}.
\end{itemize}

\begin{table}
 \caption[]{The parameters of the Schechter function  from
     2dFGRS , SDSS and MGC . }
 \label{parameters}
 \[
 \begin{array}{lccc}
 \hline
 \hline
 \noalign{\smallskip}
parameter            & 2dFGRS                        & SDSS~(r^*)~ band  & MGC  \\ \noalign{\smallskip}
M^* ~ [mags]         &   -19.79 \pm 0.04             & -20.44 \pm 0.01 &  -19.60 \pm 0.04 \\ \noalign{\smallskip}
\alpha               &   -1.19  \pm 0.01             & -1.05  \pm 0.01 &  -1.13 \pm 0.02 \\ \noalign{\smallskip}
\Phi^* ~[h~Mpc^{-3}] &   (1.59   \pm 0.1)10^{-2}     &(1.49   \pm 0.04)10^{-2}      &  (1.77   \pm 0.15)10^{-2} \\ \noalign{\smallskip}
 \hline
 \hline
 \end{array}
 \]
 \end {table}

Over the years many modifications  have been  made
to the standard Schechter function in order   to improve its fit:
we report three of them.
When the fit  of the rich clusters  luminosity function
is not satisfactory
a two-component Schechter-like function is introduced
, see~\cite{Driver1996}.
%
This two-component function is the Schechter function  when
 $ L_{Dwarf} <    L  <  L_{max} $
 and has $(\frac{L}{L_{Dwarf}})^{\alpha_{Dwarf}}$ dependence
  when $ L_{min} <   L  < L_{Dwarf} $: $L_{Dwarf}$ represents  the magnitude where
dwarfs first dominate over giants ,  ${\alpha_{Dwarf}}$ the faint
slope parameter for the dwarf population ,  the index {\it min}
and {\it max} denote the minimum and the maximum.

Another example is
the hybrid Schechter+power-law fit
to fit the faint-end of the K-band, see~\cite{Bell2003}.

Another function  introduced in order to fit  the case of
extremely low luminosity  galaxies is the  double  Schechter
function  , see~\citet{Blanton_2005} , where  the parameters
$\Phi^*$ and $\alpha$ that characterise the    Schechter function
have  been doubled in $\phi_{\ast,1},\phi_{\ast,2}$ and
$\alpha_1,\alpha_2$.
The previous efforts bring the attention toward 
two ranges in luminosity for galaxies  : an overall zone 
from high luminosity to low luminosity and the 
low luminosity zone.
This situation remembers the case  of the stars in which three zones
are considered , see  \cite{Scalo1986}, 
\cite{Kroupa1993}, and \cite{Binney1998} ;
in this case the range of existence
of the  zones  as well as the exponent that 
characterises the power law behaviour
are functions of the   investigated environment.
These  three  zones in the mass distribution of the stars 
 have  been investigated
 at the light of the physical processes 
 in \cite{Elmegreen2004}; they correspond  to
 brown dwarf masses $\approx~0.02\mathcal {M}_{\sun}$,
 to intermediate mass stars and high mass stars.
%

The starting point of this paper is a statistical distribution in
the mass of the galaxies , $\mathcal { M }$ as given by a a
standard gamma variate  with a   range of existence $ 0 <  \mathcal
{ M } <  \infty $     . This distribution  describes the area of
the irregular Voronoi polygons. This distribution in mass can be
converted in a new statistical distribution for the luminosity of
galaxies through an analogy with the physics of the stars. This
new distribution in luminosity ,$L$ ,is characterised by a  range of
existence $ 0 <   { L } <  \infty $  and a local maximum , named
mode. A second distribution in the masses starts from a truncated
Pareto distribution , after~\cite{Pareto} , with a range of
existence $ \mathcal {M}_{min} <  \mathcal { M } < \mathcal
{M}_{max}  $  . The standard procedure of conversion from from
mass to luminosity allows us to derive a truncated Pareto type
luminosity function that has a range of existence $ L_{min} < { L }
< L_{max} $  . In this distribution the mode is at $L_{min}$.

Section~\ref{luminosity_my}
first introduces the 2D Voronoi diagrams
and then    describes
the mathematical details
that  allow  us  to deduce two new physical functions
for luminosity of galaxies  and
Section~\ref{test} reports a first test
based on the SDSS photometric catalog.

In Section~\ref{secz}   the red-shift dependence
of the   Schechter function  and  the
first new  function   are explored in  detail.
Section~\ref{secmass} reports  the mass
evaluation for galaxies  based
on the first luminosity function as well as a new formula
for the limiting  mass.
Section~\ref{conclusions} summarises the results.

\section{From the mass  to the magnitude}
The first paragraph briefly introduces the 2D Voronoi diagrams as
produced by two  types of seeds. These   two statistical
distributions adopted to fit the area
 of the irregular Voronoi
polygons   can be taken as a starting point to construct two
luminosity functions for galaxies.
\subsection{The Voronoi Diagrams}
When  the seeds  are randomly and
uncorrelated distributed, are
called Poisson Voronoi diagrams.
A great number of natural phenomena are
described by Poisson Voronoi diagrams , we cite some of them  :
lattices in quantum field theory ,
see  \cite{Drouffe1984} ;
 conductivity and
percolation  in granular composites , 
see  \cite{Jerauld1984_a} and  \cite{Jerauld1984_b};
 modelling growth of metal clusters on amorphous
substrates , see  \cite{Dicenzo1989};
the statistical mechanics of simple glass 
forming systems in 2D  , see \cite{Hentschel2007};
modelling of material interface evolution 
in grain growth of polycrystalline
materials , see ~\cite{Lee2006}.
When  the seeds  are randomly and
uncorrelated distributed, are
called Poisson Voronoi diagrams.
A great number of natural phenomena are
described by Poisson Voronoi diagrams , we cite some of them  :
lattices in quantum field theory ,
see  \cite{Drouffe1984} ;
 conductivity and
percolation  in granular composites , 
see  \cite{Jerauld1984_a} and  \cite{Jerauld1984_b};
 modelling growth of metal clusters on amorphous
substrates , see  \cite{Dicenzo1989};
the statistical mechanics of simple glass 
forming systems in 2D  , see \cite{Hentschel2007};
modelling of material interface evolution 
in grain growth of polycrystalline
materials , see ~\cite{Lee2006}.
A review of the Voronoi diagrams applied 
to the spatial distribution of the galaxies
can be found in~\cite{Zaninetti2006}.
Here we are interested in the fragmentation of a 2D
layer of thickness negligible in respect
to the main dimension.
A typical dimension of the layer can be found  as follows.
The averaged observed diameter of the galaxies is
\begin{equation}
\overline{D^{obs}} \approx  0.6  {D_{max}^{obs}} = 2700 \frac{Km}{sec}
= 27~Mpc
\quad ,
\label{dobserved}
\end {equation}
where $D_{max}^{obs}=4500~\frac{Km}{sec}$  corresponds to the
extension of the maximum void visible on the CFA2 slices. In the
framework of the theory of the primordial explosions
,see~\citet{charlton} and \citet{ferraro}, this means that the
mean observed area of a bubble ,$\overline{A^{obs}} $, is
\begin{equation}
\overline{A^{obs}} \approx  4 \pi (\frac{D_{max}^{obs}}{2})^2
=2290 Mpc^2
\quad .
\label{aobserved}
\end {equation}
The averaged area of a face of  Voronoi polyhedra ,
$\overline{A_V} $, is
\begin{equation}
\overline{A_V}  = \frac{\overline{A^{obs}}} {\overline{N_F}}
\quad ,
\end {equation}
where  ${\overline{N_F}}$ is the averaged number of irregular
faces of the Voronoi polyhedra,
i.e.  ${\overline{N_F}}$=16 , see~\citet{okabe,Zaninetti2006}.
The averaged side of a face of a irregular polyhedron , $L_V$ ,
is
\begin{equation}
\overline{L_V}  \approx  \sqrt {\overline{A^{obs}}} \approx 12~Mpc
\quad .
\end {equation}
The thickness of the layer  , $\delta$ ,
can be derived from the shock theory
, see~\cite{deeming}, and is  1/12 of the radius
of the advancing shock ,
 \begin{equation}
\delta = \frac { {D_{max}^{obs}}  } {2 \times 12 } \approx 1.12 Mpc
\quad .
\end{equation}
The number of galaxies in this typical layer , $N_G$, can be found
by 
multiplying $n_*\approx 0.1$ , the density of galaxies , for the
volume of the cube of side 12~$Mpc$ : i.e. $N_G \approx 172 $.

\begin{figure}
\begin{center}
\plotone{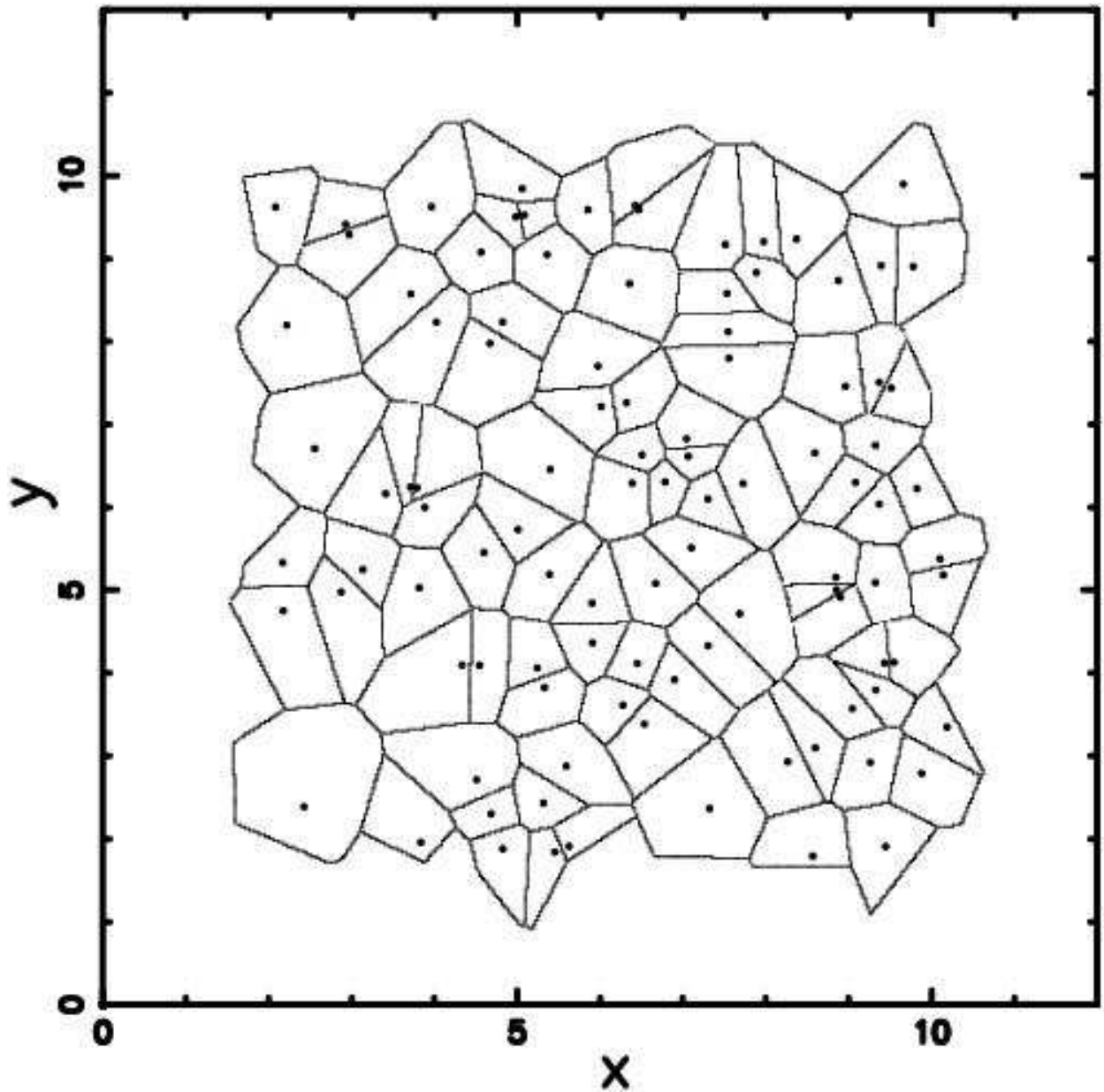}
\end  {center}
\caption {
The Voronoi--diagram in 2D when random seeds are used.
The selected region  comprises  77 seeds and
$c=5.1$ .
}
\label{area2D_random}
\end{figure}

The more common way to insert the seeds of the Voronoi polygons
is a random sequence in the $X$  and $Y$ direction ,
see Figure~\ref{area2D_random}.

The distribution of the area of the irregular Voronoi polygons is
fitted with a Kiang  function, see formula~(\ref{kiang}) in
Appendix~\ref{appendixa},:
\begin{equation}
 H (x ;c ) = \frac {c} {\Gamma (c)} (cx )^{c-1} \exp(-cx)
\quad ,
\label{kianglum}
\end{equation}
and the captions of
Figure~\ref{area2D_random} also report the value of $c$
as deduced from the parameters of the sample.

Another way to insert the seeds is through a
truncated Pareto distribution , see Appendix~\ref{appendixb};
this  is an example of non-Poissonian seeds.  

\begin{figure}
\begin{center}
\plotone{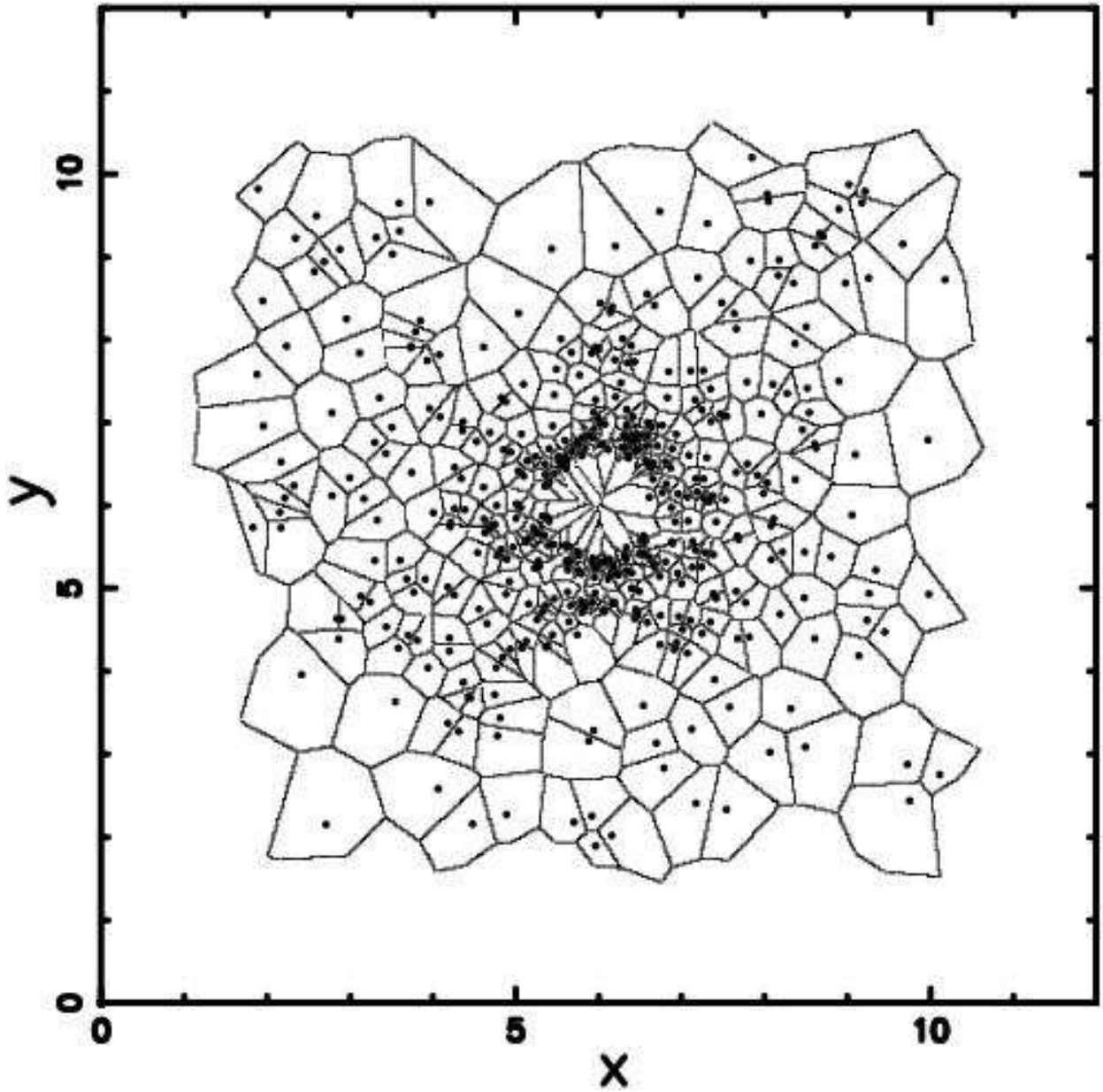}
\end  {center}
\caption {
The Voronoi--diagram in 2D when truncated Pareto  seeds are used.
The selected region  comprises 124  seeds and
$d=0.27$ as deduced from MLE ,
 see Appendix~\ref{appendixb}   .
}
\label{area2D_pareto}
\end{figure}

In polar coordinates the radial distribution of seeds
$ p(r) $
will vary  according to 
\begin{equation}
p(r) \propto \frac {1}{r^{d+1}}
\quad ,
\end {equation}
where  $r$ is the distance from the center of the box
,
see Figure~\ref{area2D_pareto}.

\subsection{A first main physical luminosity  distribution}
\label{luminosity_my}

In order to start,   we briefly review   how a
probability density function  ( in the following pdf)  $f(x)$ changes to 
$g(y)$  when a new variable $y (x)$  is introduced~. We limit  ourselves
to the case in which $y(x)$ is a unique transformation.
The rule for
transforming  a pdf is
\begin{equation}
g(y) =  \frac  {f(x) } {\vert\frac {dy } {dx} \vert} \label{trans}
\quad.
\end{equation}
We start by assuming that the  masses  of the galaxies are distributed like
a Kiang  function, see formula~(\ref{kiang}) in  Appendix~\ref{appendixa}
or the previous paragraph.
This assumption  is  justified by the fact  that
the various processes that lead to the formation
of a galaxy  can follow a  random fragmentation in 2D.
The first
transformation is
\begin{equation}
x = \frac  {\mathcal M} {\mathcal M^*} \label {first} \quad  ,
\end{equation}
and therefore  equation~(\ref{kiang}) changes to
\begin{equation}
\Psi ({\mathcal M }) d{\mathcal M }
      = \frac {\left ( \frac  {\mathcal M }{\mathcal M^*} \right )^ {c-1}
      e ^{-\frac {\mathcal M}{\mathcal M^*}} } {\Gamma (c)}
      d\frac  {\mathcal M }{\mathcal M^*}
\label{psi} \quad.
\end {equation}

This is a gamma distribution with a scale parameter ${\mathcal M^*}$
and a shape parameter {\it c},
and its averaged value
is
\begin{equation}
{\langle{\mathcal M}\rangle}=c {\mathcal M^*}
\quad .
\label{massamedia}
\end{equation}

The mass-luminosity relationship in the case of the stars
is well established both from a theoretical point of view,
$L  \propto M^3 $ or $L  \propto M^4 $  ,see ~\cite{lang},   and
 from an observational point of view , $L  \propto M^{3.43} $
in the case of MAIN,V  , see~\cite{zaninetti05} for further
details.
A power law which  is introduced by analogy
regulates the relationship between mass and luminosity
of galaxies, but in this case the regulating parameter $a$
does not have a theoretical counterpart.
The second  transformation  is
\begin{equation}
\frac  {\mathcal M }{\mathcal M^*}  = \left ( \frac  {L  }{L^*}
\right ) ^{\frac{1}{a} } \quad  ,
\label{massa}
\end{equation}
where  1/a is  an exponent  that connects the mass to the
luminosity. The pdf (\ref{psi} ) is therefore transformed  into
 the
following:
\begin{eqnarray}
\Psi (L) dL  =&&  (\frac{1}{a \Gamma(c) } ) (\frac {\Psi^*}{L^*})
\left (\frac {L}{L^*} \right  )^{\frac{c-a}{a}} \nonumber\\
&&\times \exp \left  ( {-
\left ( \frac {L}{L^*}\right  )^{\frac{1}{a}}} \right   ) dL \quad
, \label{equation_schechter_mia}
\end {eqnarray}
where $\Psi^*$  is  a normalisation factor which defines the
overall density of galaxies , a  number  per cubic $Mpc$.
The mathematical range of existence is  $  0 \leq L < \infty $
; conversely   the astronomical range is  $  L_{\mathrm{min}}  \leq L <
L_{\mathrm{max}} $.

The relationship connecting the absolute magnitude, $M$ ,
 of a
galaxy  with its luminosity is
\begin{equation}
\frac {L}{L_{\sun}} =
10^{0.4(M_{bol,\sun} - M)}
\quad ,
\label{mlrelation}
\end {equation}
where $M_{bol,\sun}$ is the bolometric luminosity
of the sun , which   according to \cite{cox}
is $M_{bol,\sun}$=4.74.

The third
and last transformation connects the luminosity with the absolute
magnitude
\begin{eqnarray}
\Psi (M) dM  = && (0.4  ln 10 \frac {1}{a \Gamma(c)}) \Psi^*
10^{0.4(\frac{c}{a}) (M^*-M)} \nonumber \\
&& \times \exp \bigl ({-
10^{0.4(M^*-M)(\frac{1}{a})}} \bigr)  dM \quad.
\label{equation_mia}
\end {eqnarray}
This data oriented function contains the  parameters $M^*$ ,{\it a},
{\it c}
and  $\Psi^*$ which  can be derived from the operation of fitting
the observational data.
Other  interesting quantities are  the  mean
luminosity per unit volume,
$j$ ,
\begin{equation}
\label{eqnj}
j = \int ^{\infty}_0 L \Psi(L) dl = L^* \Psi^* \frac {\Gamma
(c +a)} {\Gamma (c)}  \quad ,
\end{equation}
and the averaged luminosity ,${ \langle L \rangle }$ ,
\begin{equation}
\label{lmedia}
\langle L \rangle =\frac{j}{\Psi^*}= L^*  \frac {\Gamma
(c +a)} {\Gamma (c)}
 \quad.
\end{equation}

The  density of  galaxies is
\begin{equation}
n_* = \frac {j}{L^*}
\quad ,
\end{equation}
and  the mean separation between galaxies ,
\begin{equation}
d_* = n_*^{- 1/3}
\quad.
\end{equation}
The symbols $ j$, $n_*$ and $d_*$ are introduced  as in \citet{pad}.

Another way to compute the  density of  galaxies , now $n_{**}$,
of  the  ${\mathcal M}-L$ function is
\begin{equation}
n_{**} = \Psi^*
\quad .
\end{equation}
The position of the maximum in magnitudes is at
\begin{equation}
M_{p,max} =
{\it M^*}
- 1.085\,\ln  \left( c \right) a
\quad .
\end{equation}

\subsection{The luminosity  distribution for low luminosity galaxies}
The Pareto distribution 
can  model
nonnegative data with a power law probability tail. 
In many practical applications, it is natural to consider an 
upper bound that 
truncates the tail  \cite{Cohen1988,Devoto1998,Aban2006}.
The truncated Pareto distribution has a wide range of applications ,
we list some of them :
data analysis~\cite{Aban2006}and  \cite{Rehfeldt1992};
forest  fire area in the Australian Capital Territory,
fault offsets in the
Vernejoul coal field,
hydrocarbon volumes in the Frio Strand Plain
exploration play and fault lengths on Venus, see \cite{Burroughs2001}.

In the case of the stars ,  the low mass
distribution of  masses , see~\cite{Salpeter1955}, can be represented
by   a law of the type
$p  ( {\mathcal{M_S}}) \propto   {\mathcal{M_S}}^{-2.35}$,
where $p   ( {\mathcal{M_S}})$ represents  the probability
of having  a mass between $ {\mathcal{M_S}}$ and
$ {\mathcal{M_S}}+d{\mathcal{M}}$~.
By analogy we introduce a truncated Pareto distribution,
see Appendix~\ref{appendixb} ,
for the mass of galaxies
\begin{equation}
\Psi_{LL} ({\mathcal M }) d{\mathcal M }
      = \frac {C}
              {
              (\frac {\mathcal M }{\mathcal M^*})^{d+1}
              }
      d (\frac  {\mathcal M }{\mathcal M^*})
\label{psill} \quad ,
\end {equation}
where the index $LL$ stands  for Low Luminosity
and the range of existence is
 $   {\mathcal M_{min} } \leq  {\mathcal M } \leq  {\mathcal M_{max} }  $
where $min$ and $max$ denote the minimum and maximum mass.
Once the constant $C$ is computed as in
Appendix~\ref{appendixb} we obtain
\begin{equation}
\Psi_{LL} ({\mathcal M }) d{\mathcal M }
      =
\frac{d }
{
\left(  \left( {\frac {{\it {\mathcal M_{min} }}}{{\it {\mathcal M^*}}}} \right) ^{-d}- \left(
{\frac {{\it {\mathcal M_{max} }}}{{\it {\mathcal M^*}}}} \right) ^{-d} \right)
 \left( {\frac { {\mathcal M }}{{\it {\mathcal M^*}}}} \right) ^{d+1}
}
      d (\frac  {\mathcal M }{\mathcal M^*})
\label{psilldue} \quad .
\end {equation}
Exactly as in the previous case we introduce the
transformation  represented by
equation~(\ref{massa})
that  connects the mass  with   the luminosity and
the distribution in luminosity is
\begin{equation}
\Psi_{LL} (L) dL
=\Psi_{LL}^*
\frac
{
d \left( {\frac {L}{{\it L^*}}} \right) ^{-{\frac {d+a}{a}}}
}
{
\left(  \left( {\frac {{\it L_{min}}}{{\it L^*}}} \right) ^{-{\frac
{d}{a}}}- \left( {\frac {{\it L_{max}}}{{\it L^*}}} \right) ^{-{\frac {
d}{a}}} \right) a
}
d (\frac { L}{L^*})
\quad ,
\label{lum_pareto}
\end{equation}
with  the range of existence as
 $   {L_{min} } \leq  { L } \leq  {L_{max} }  $
and $\Psi_{LL}^*$ representing the normalisation.
The  mean
luminosity per unit volume,
$j$ ,
\begin{equation}
\label{eqnjpareto}
j = \int ^{L_{max}}_{L_{min}} L \Psi_{LL} (L) dl =
\Psi_{LL}^*
\frac
{
d \left( -{{\it {L_{max}}}}^{2} \left( {\frac {{\it {L_{max}}}}{{\it L^*}}}
 \right) ^{-{\frac {d+a}{a}}}+{{\it {L_{min}}}}^{2} \left( {\frac {{\it
{L_{min}}}}{{\it L^*}}} \right) ^{-{\frac {d+a}{a}}} \right)
}
{
\left( d-a \right)  \left(  \left( {\frac {{\it {L_{min}}}}{{\it L^*}}}
\right) ^{-{\frac {d}{a}}}- \left( {\frac {{\it {L_{max}}}}{{\it L^*}}}
 \right) ^{-{\frac {d}{a}}} \right) {\it L^*}
}
 \quad.
\end{equation}
The  distribution in magnitude is
\begin{equation}
\Psi_{LL} (M) dM  =\Psi^*_{LL}
\frac
{
 0.4\,d{10}^{- 0.4\,{\frac {d \left( {\it
M^*}- \,{\it am} \right) }{a}}}\ln  \left( 10 \right)
}
{
\left( {10}^{- 0.4\,{\frac { \left( {\it M^*}- \,{\it {M_{max}}}
\right) d}{a}}}- \,{10}^{- 0.4\,{\frac { \left( {
\it M^*}- \,{\it {M_{min}}} \right) d}{a}}} \right) a
}
dM \quad ,
\label{equation_miapareto}
\end {equation}
with  the range of existence as
 $   {M_{min} } \leq  { M } \leq  {M_{max} }  $.
This  distribution in magnitude  contains the parameters ${M_{min}
}$ and ${M_{max} }$ which  are the minimum and maximum magnitude  of
the considered catalog and the parameters $a$ , $d$ , and
$\Psi_{LL}$ which  are derived from the fitting of the data.

\section{Application to a real sample of galaxies }
\label{test}

The data of the luminosity function for galaxies 
in five bands of the  SDSS are available at  
http://cosmo.nyu.edu/blanton/lf.html 
and are  discussed from an astronomical point view 
in~\citet{blanton}.

The analysis of the new luminosity function 
was split  in two.
The case  from high luminosities 
to low luminosities was fitted by  
$\Psi (M) $ , equation~(\ref{equation_mia}) .
The data have   been  processed
through the  Levenberg--Marquardt  method ( subroutine
MRQMIN in \citet{press})
in order
to find the three parameters   {\it a},  $M^*$ , $\Psi^*$
;  {\it c} conversely  is introduced by hand.
In order  to associate a statistical probability 
to each fit we have chosen a range in magnitude 
such as $M< M_{max}$ where  $M_{max}$ represents 
the selected maximum magnitude of the sample.

The results are reported in Table~\ref{para_physical_SDSS_data} 
together with  the derived quantities  
$ j$, $n_*$ ,  $d_*$ and their   uncertainties.
Table~\ref{para_physical_SDSS_data} also reports 
$M_{max}$ ,
the number of elements  $N$ belonging to the sample ,
the  merit function $\chi^2$ and
 the associated $p$--value
 that has   to be understood as the
 maximum probability to obtain a better fitting, 
 see formula~(15.2.12) in \citet{press}:
\begin{equation}
p=1- GAMMQ (\frac{N-3}{2},\frac{\chi^2}{2} ) 
\quad ,
\end{equation}
where GAMMQ is a subroutine  for the incomplete gamma function.
\clearpage
\begin{deluxetable}{crrrrr}
\tabletypesize{\scriptsize}
\rotate
\tablecaption{Parameters of Fits to Luminosity 
          Function    in  SDSS Galaxies through the 
          $\mathcal {M}-L$  function .\label{para_physical_SDSS_data} }
\tablewidth{0pt}
\tablehead{
\colhead{Band } & 
\colhead{$u^*$} & 
\colhead{$g^*$} & 
\colhead{$r^*$}  & 
\colhead{$i^*$} &
\colhead{$z^*$} 
}
\startdata
c          &  1.1  $\pm$ 0.2                               
           &  1.0  $\pm$ 0.2               
           &  1.1    $\pm$ 0.2      
           &  2    $\pm$ 0.2
           &  1.7    $\pm$ 0.2
                         \\ 
$M^*$ $[mags]$     &   -16.58 $\pm$ 0.018         
                   &   -18.29 $\pm$ 0.008   
                   &   -18.77 $\pm$ 0.007        
                   &   -18.26  $\pm$ 0.01   
                   &   -18.79 $\pm$ 0.004 
                          \\
$\Psi^*$ $[h~Mpc^{-3}]$   &   0.069   $\pm$ 0.001 
                          &   0.043   $\pm$ 0.0003   
                          &   0.043   $\pm$ 0.000       
                          &   0.032   $\pm$ 0.0002   
                          &   0.034   $\pm$ 0.003
                          \\
a          &   1.40   $\pm$ 0.007               
           &   1.32   $\pm$ 0.003    
           &   1.5   $\pm$ 0.002         
           &   1.74   $\pm$ 0.003   
           &   1.70   $\pm$ 0.014   
                          \\ 
j   [mags]       &   1.40 $L^* \Psi^*$           
                 &   1.18 $L^* \Psi^*$    
                 &   1.50 $L^* \Psi^*$       
                 &   4.39 $L^*  \Psi^*$    
                 &   3.2  $L^*  \Psi^*$    
                           \\                    
$n_*$ [$Mpc^{-3}]$    &  0.097      $\pm$ 0.022
                      &  0.051      $\pm$ 0.012   
                      &  0.066      $\pm$ 0.016         
                      &  0.14      $\pm$ 0.02   
                      &  0.11      $\pm$ 0.04
                           \\                   
d  [Mpc]              &  2.17      $\pm$ 0.16     
                      &  2.67      $\pm$ 0.21
                      &  2.47     $\pm$ 0.19
                      &  1.91      $\pm$ 0.09   
                      &  2.06      $\pm$ 0.14    
                           \\                   
$M_{max}$ $[mags]$    &   -15.78 
                      &   -18.2 
                      &   -19  
                      &   -19.3
                      &   -20 
                           \\
N                            & 483
                             & 404 
                             & 400
                             & 471 
                             & 442 
                                 \\
$\chi^2$   &    321                               
           &    386                
           &    233
           &    325     
           &    649
                           \\                    
$p=1- GAMMQ(\frac{N-3}{2},\frac{\chi^2}{2} ) $&  0
                             & 0.31
                             & 0 
                             & 1.19 $10^{-7}$
                             & 1.0 
\\
\enddata
\end{deluxetable}

The  uncertainties  are   found by implementing the error
propagation equation (often called law of errors of Gauss).
The low luminosities range conversely was fitted
through  $\Psi_{LL} (M)$ , equation~(\ref{equation_miapareto}) and
the results are reported in Table~\ref{para_physical_SDSS_LL}.

\clearpage
\begin{deluxetable}{crrrrr}
\tabletypesize{\scriptsize}
\rotate
\tablecaption{Parameters of Fits to  Luminosity 
          Function    in  SDSS Low Luminosities Galaxies through the 
          $\Psi_{LL}$ function~.
          \label{para_physical_SDSS_LL} }
\tablewidth{0pt}
\tablehead{
\colhead{Band } & 
\colhead{$g^*$} & 
\colhead{$r^*$}  & 
\colhead{$i^*$} &
\colhead{$z^*$} 
}
\startdata
d          &  0.3    $\pm$ 0.1                               
           &  0.4    $\pm$ 0.1
           &  0.5    $\pm$ 0.1
           &  0.9    $\pm$ 0.1
                         \\ 

$M^*$ $[mags]$     &   -17.2   $\pm$ 0.1         
                   &   -18.8   $\pm$ 0.1
                   &   -17.39  $\pm$ 0.1   
                   &   -19.3   $\pm$ 0.1  
                          \\
$\Psi^*_{LL}$ $[h~Mpc^{-3}]$   &   0.043   $\pm$ 0.0043
                          &   0.040   $\pm$ 0.0040       
                          &   0.026   $\pm$ 0.0032   
                          &   0.035   $\pm$ 0.003
                          \\
a          &   2.2   $\pm$ 0.1               
           &   1.3   $\pm$ 0.1         
           &   2.7   $\pm$ 0.1   
           &   2.3   $\pm$ 0.1
                          \\ 
range [mags]          &   -18.2 $\leq$ M $\leq$ 16.33 
                      &   -19.0 $\leq$ M $\leq$ 16.31 
                      &   -19.3 $\leq$ M $\leq$ 17.22 
                      &   -20.0 $\leq$ M $\leq$ 17.48  
                           \\
N                            & 194
                             & 273
                             & 237 
                             & 297 
                                 \\
$\chi^2$   &    204                              
           &    379
           &    476     
           &    313
                           \\                    
\enddata
\end{deluxetable}

The Schechter function, conversely fits all the range 
in luminosities and 
Table~\ref{para_Schechter_SDSS_data}  reports  
the data  that come  out from the fitting procedure.
Table~\ref{para_Schechter_SDSS_data}  also reports
$M_{p,max} $  , the value in magnitude where the 
Schechter  function peaks ; this value is defined
when  $\alpha \> -1 $ , otherwise 
we leave the box  blank.
\clearpage
\begin{deluxetable}{crrrrr}
\tabletypesize{\scriptsize}
\rotate
\tablecaption{Parameters of Fits to Luminosity 
          Function    in SDSS  through the 
          Schechter  function~.
         \label{para_Schechter_SDSS_data}}
\tablewidth{0pt}
\tablehead{
\colhead{Band } & 
\colhead{$u^*$} & 
\colhead{$g^*$} & 
\colhead{$r^*$}  & 
\colhead{$i^*$} &
\colhead{$z^*$} 
}
\startdata
$\alpha$   
&  -0.90 $\pm$ 0.01
&  -0.88 $\pm$ 0.007
&  -1.04 $\pm$ 0.004          
&  -0.99 $\pm$ 0.005
&  -1.07 $\pm$ 0.02
\\ 
$M^*$ $[mags]$     
&   -17.92 $\pm$ 0.006         
&   -19.38 $\pm$ 0.004   
&   -20.43 $\pm$ 0.003
&   -20.81 $\pm$ 0.004
&   -21.18 $\pm$ 0.017 
\\
$M_{p,max} $ $[mags]$     
&   -17.92          
&   -19.38    
&   ~ 
&   -20.81
&   ~
\\
$\Phi^*$ $[h~Mpc^{-3}]$   
&   0.030  $\pm$ 0.0003 
&   0.021   $\pm$ 0.0001
&   0.015  $\pm$ 0.00008       
&   0.0147   $\pm$ 0.00008
&   0.0135   $\pm$ 0.00006
\\
j   [mags]       
&   0.95 $L^*  \Phi^*$           
&   0.92 $L^*  \Phi^*$    
&   1.02 $L^*  \Phi^*$       
&   0.99 $L^*  \Phi^*$    
&   1.04 $L^*  \Phi^*$    
\\                    
$n_*$ [$Mpc^{-3}]$    
&  0.029     $\pm$ 0.0003
&  0.02      $\pm$ 0.0001
&  0.015     $\pm$ 0.00007         
&  0.014     $\pm$ 0.00008  
&  0.014     $\pm$ 0.00007
\\                   
d  [Mpc]              
&  3.24      $\pm$ 0.013
&  3.64      $\pm$ 0.006
&  4.02      $\pm$ 0.006          
&  4.08      $\pm$ 0.008
&  4.12      $\pm$ 0.007  \\
N                            & 483
                             & 599 
                             & 674
                             & 709
                             & 740 
                                 \\

$\chi^2$   &    330                              
           &    753                
           &    2260         
           &    2282   
           &    3245
                           \\                    
$p=1- GAMMQ(\frac{N-3}{2},\frac{\chi^2}{2} ) $  & 5.96 $10^{-8}$
                             & 0.99
                             & 1.0
                             & 1.0
                             & 1.0
  \\
\enddata
\end{deluxetable}

Table~\ref{chisquare_data} reports the 
$\chi^2$ of the two zones of the new physical function,
their sum and   $\chi^2$  of the 
Schechter luminosity function.

\clearpage
\begin{deluxetable}{crrrrr}
\tabletypesize{\scriptsize}
\rotate
\tablecaption{
              Synoptic  $\chi^2$~.
          \label{chisquare_data} }
\tablewidth{0pt}
\tablehead{
\colhead{Band } & 
\colhead{$u^*$} & 
\colhead{$g^*$} & 
\colhead{$r^*$}  & 
\colhead{$i^*$} &
\colhead{$z^*$} 
}
\startdata
$\chi^2$  ~physical~luminosity~function          &   321 & 386 & 233 & 325 & 649 \\
$\chi^2$  ~luminosity~function~low~luminosities  & 0     & 204 & 379 & 476 & 313 \\
$\chi^2$  ~sum~of~two~zones                      & 321   & 590 & 612 & 801 & 962 \\
$\chi^2$  ~Schechter~luminosity~function         & 330   & 753 & 2260 & 2282 & 3245 \\
\enddata
\end{deluxetable}

\begin{figure}
   \centering
\plotone{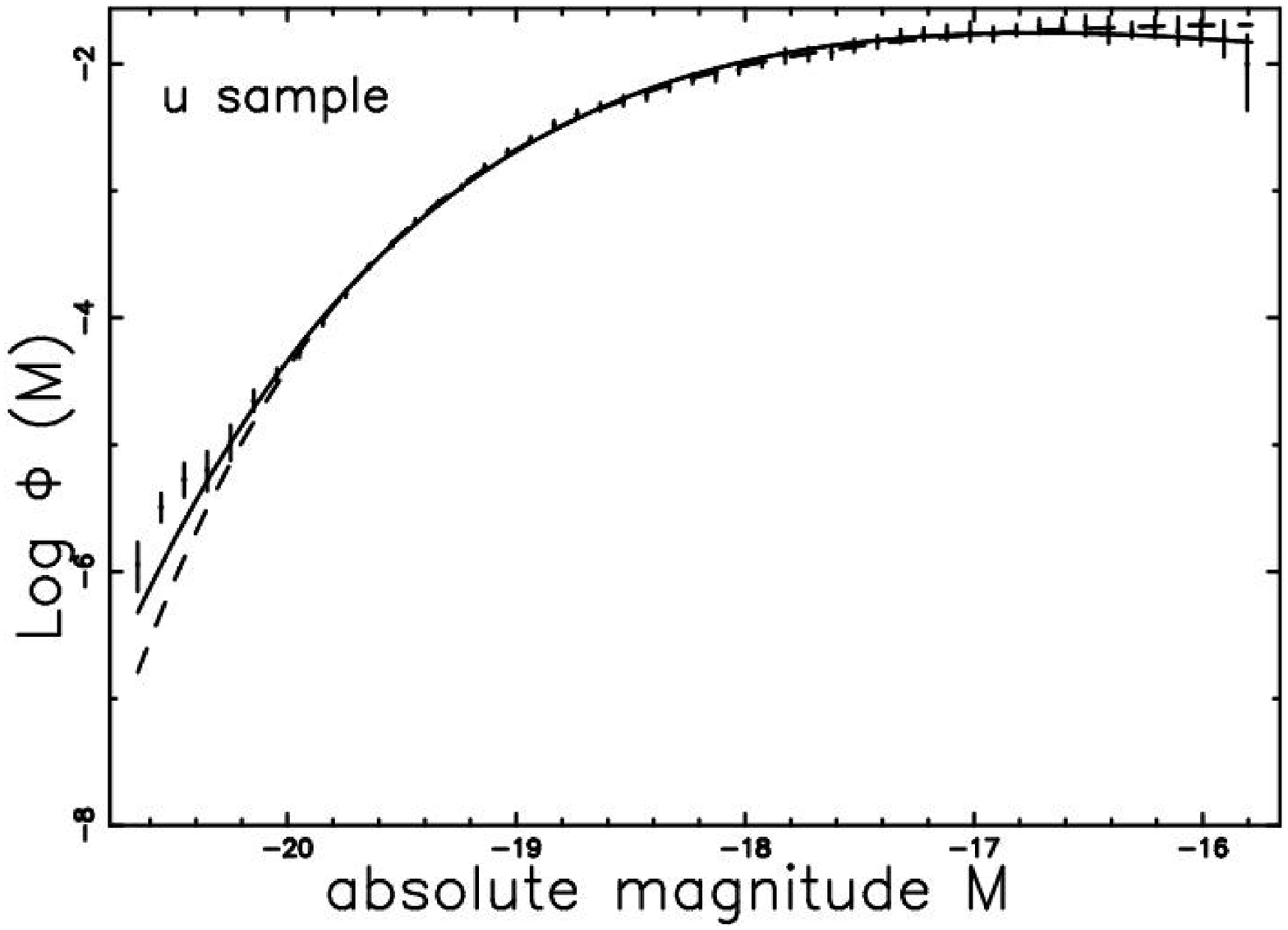}
\caption {The luminosity function  data of  
SDSS($u^*$)  are represented through the error bar. 
The fitting continuous  line represents  our  two luminosity functions
((\ref{equation_mia}) and (\ref{equation_miapareto}) )  
and the dotted 
line   represents the Schechter function.
 }
          \label{due_u} 
    \end{figure}

\begin{figure}
   \centering
\plotone{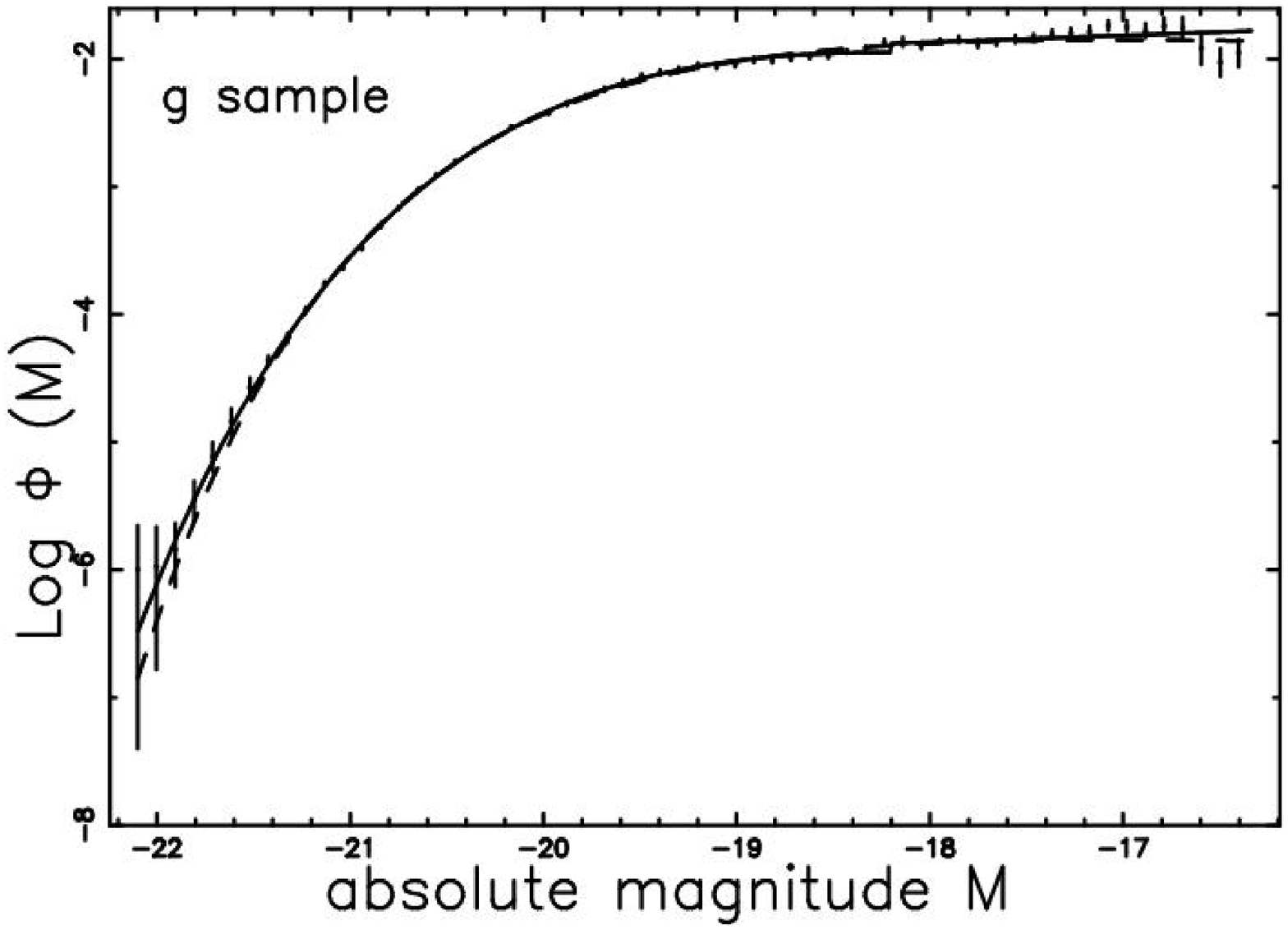}
\caption {The luminosity function  data of  
SDSS($g^*$)  are represented through the error bar. 
The fitting continuous  line represents  our  two luminosity functions 
((\ref{equation_mia}) and (\ref{equation_miapareto}) )  
and the dotted 
line   represents the Schechter function.
 }
          \label{due_g} 
    \end{figure}

\begin{figure}
   \centering
\plotone{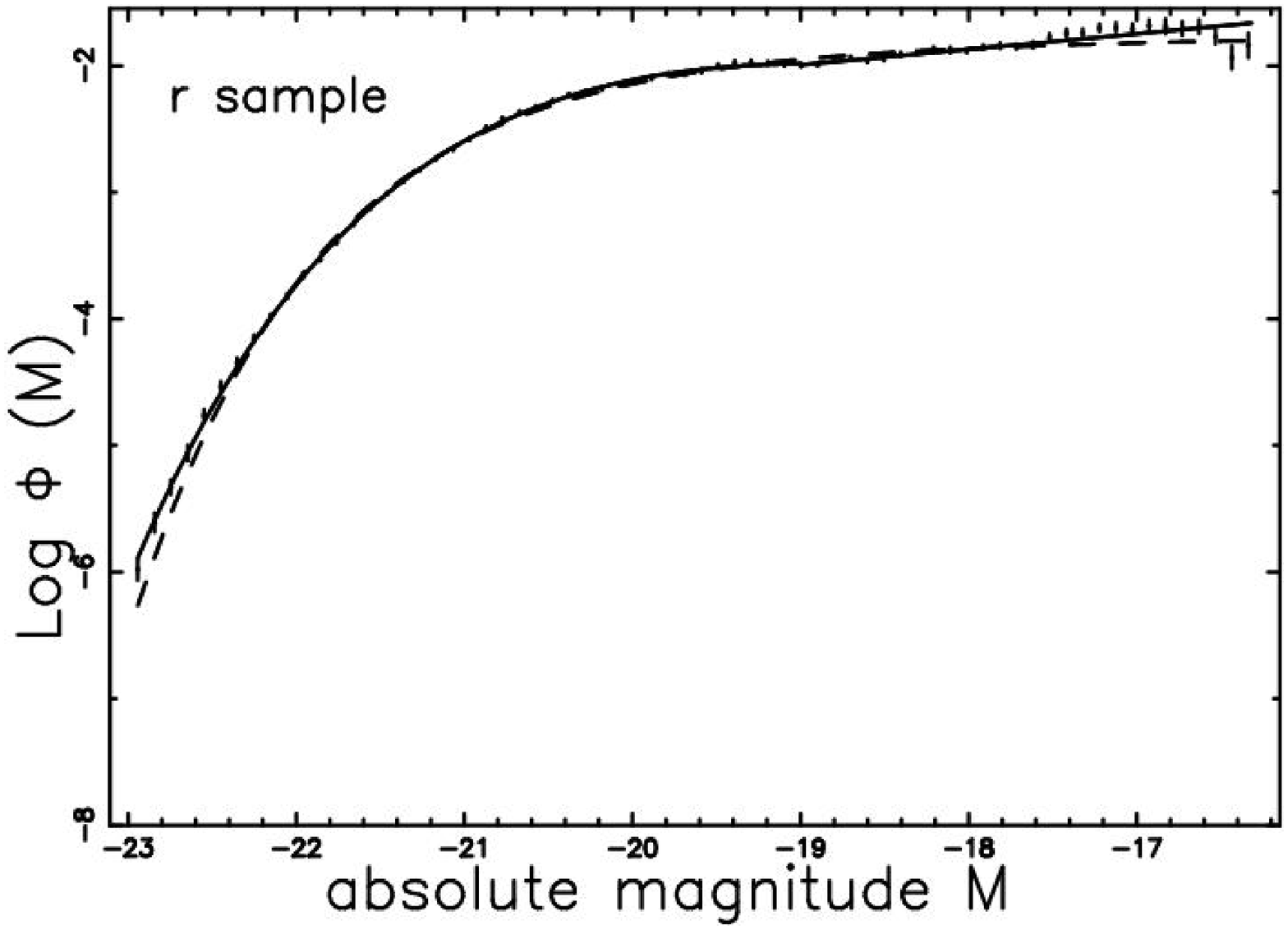}
\caption {The luminosity function  data of  
SDSS($r^*$)  are represented through the error bar. 
The fitting continuous  line represents  our  two luminosity functions 
((\ref{equation_mia}) and (\ref{equation_miapareto}) )  
and the dotted 
line   represents the Schechter function.
 }
          \label{due_r} 
    \end{figure}

\begin{figure}
   \centering
\plotone{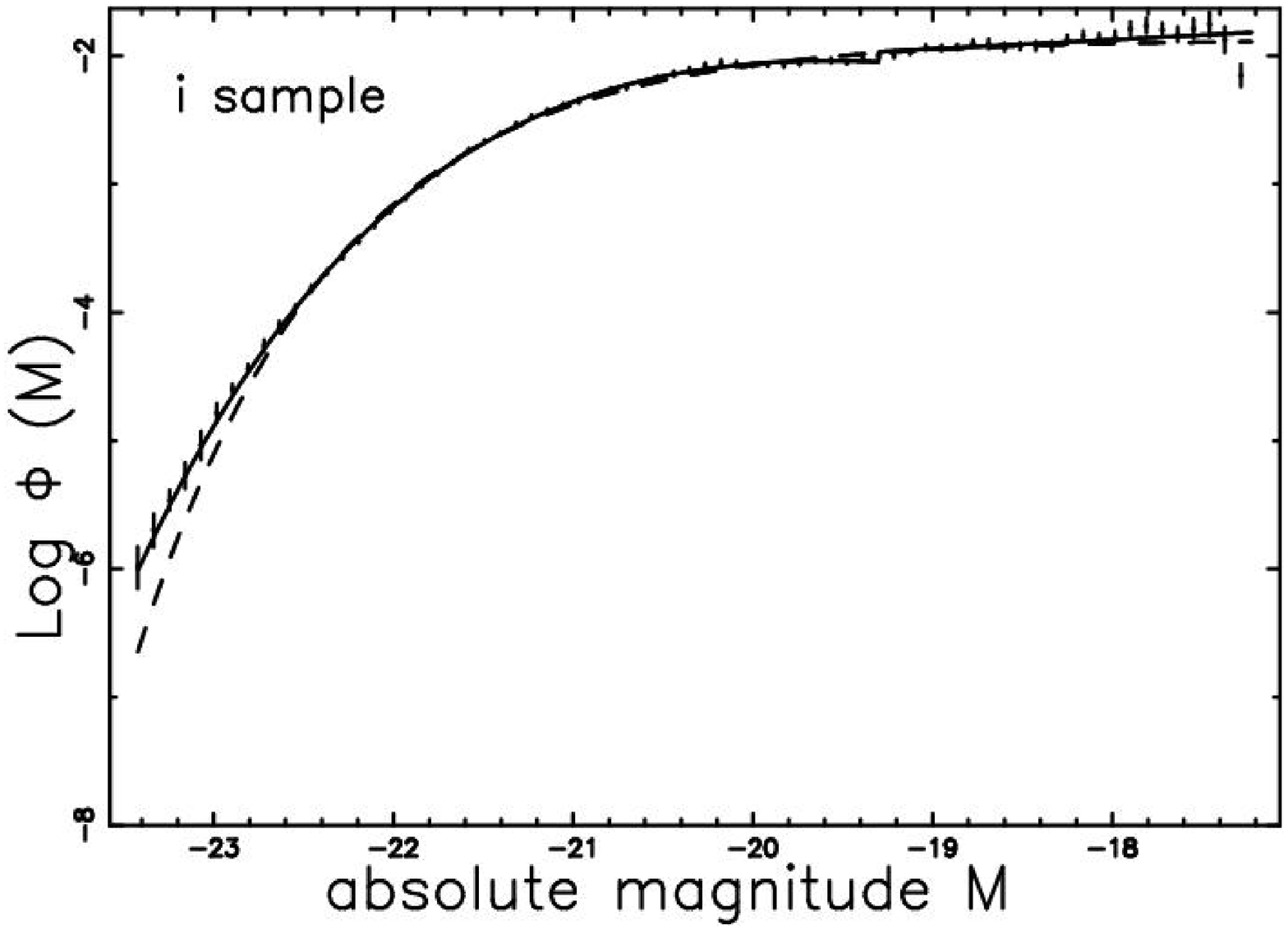}
\caption {The luminosity function  data of  
SDSS($i^*$)  are represented through the error bar. 
The fitting continuous  line represents  our  two luminosity functions 
((\ref{equation_mia}) and (\ref{equation_miapareto}) )  
and the dotted 
line   represents the Schechter function.
 }
          \label{due_i} 
    \end{figure}

\begin{figure}
   \centering
\plotone{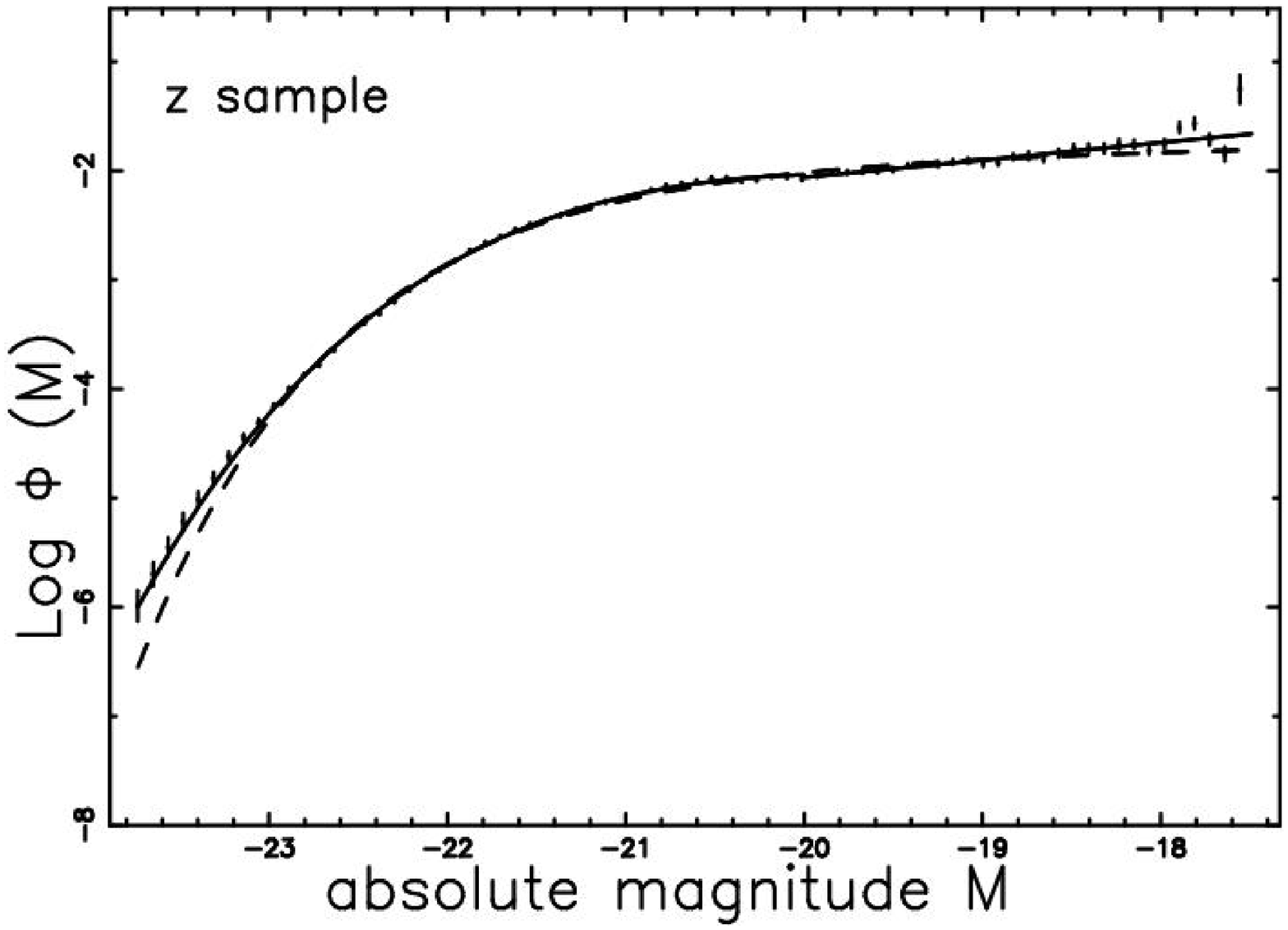}
\caption {The luminosity function  data of  
SDSS($z^*$)  are represented through the error bar. 
The fitting continuous  line represents  our  two luminosity functions 
((\ref{equation_mia}) and (\ref{equation_miapareto}) )  
and the dotted 
line   represents the Schechter function.
 }
          \label{due_z} 
    \end{figure}

\begin{figure}
   \centering
\plotone{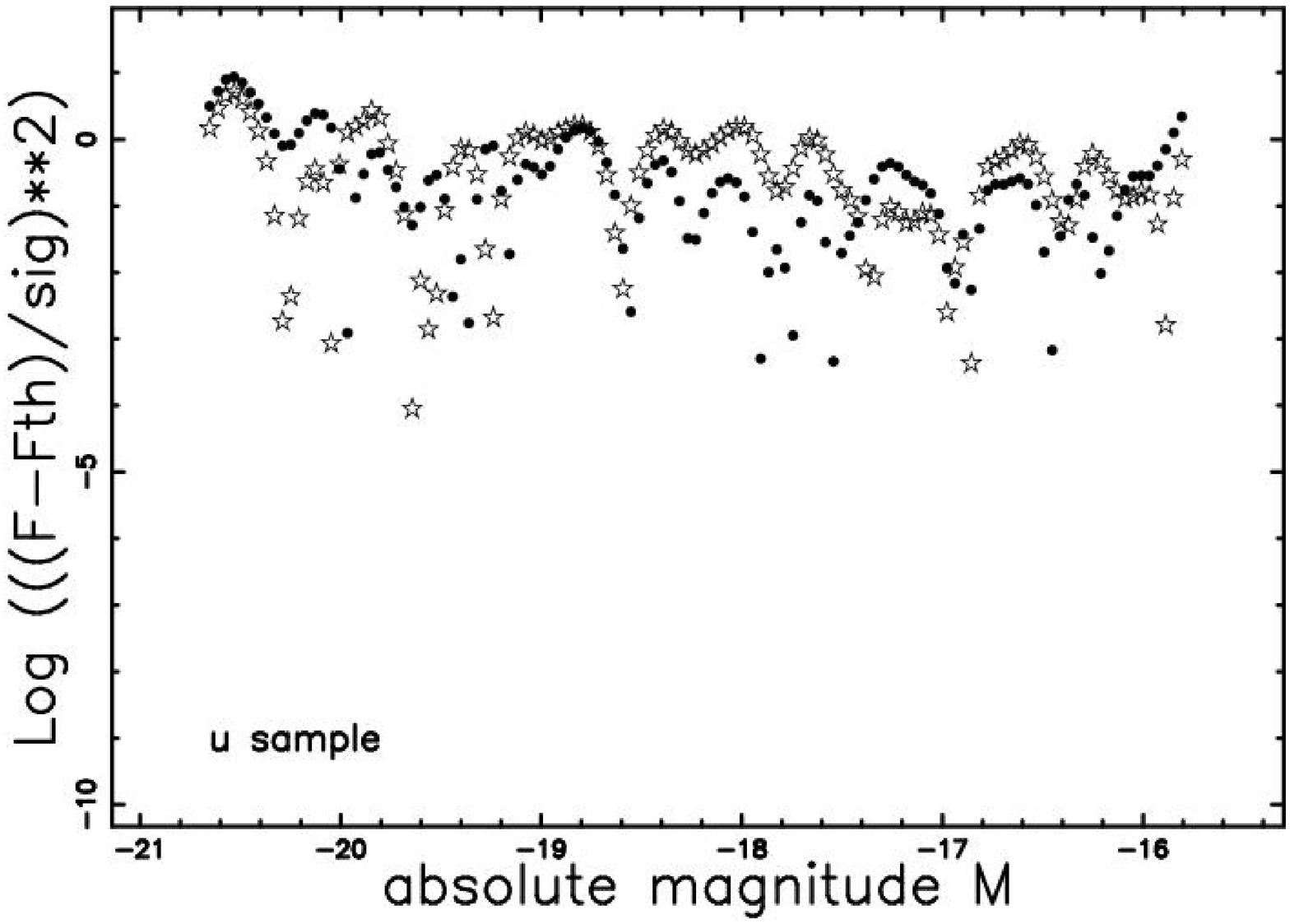}
\caption {The residuals of  the fits to 
SDSS($u^*$) data.
 The empty stars represent  our  
two luminosity functions 
((\ref{equation_mia}) and (\ref{equation_miapareto}) )  
and the filled points 
represent   the Schechter function.
 }
          \label{residui_u} 
    \end{figure}

\begin{figure}
   \centering
\plotone{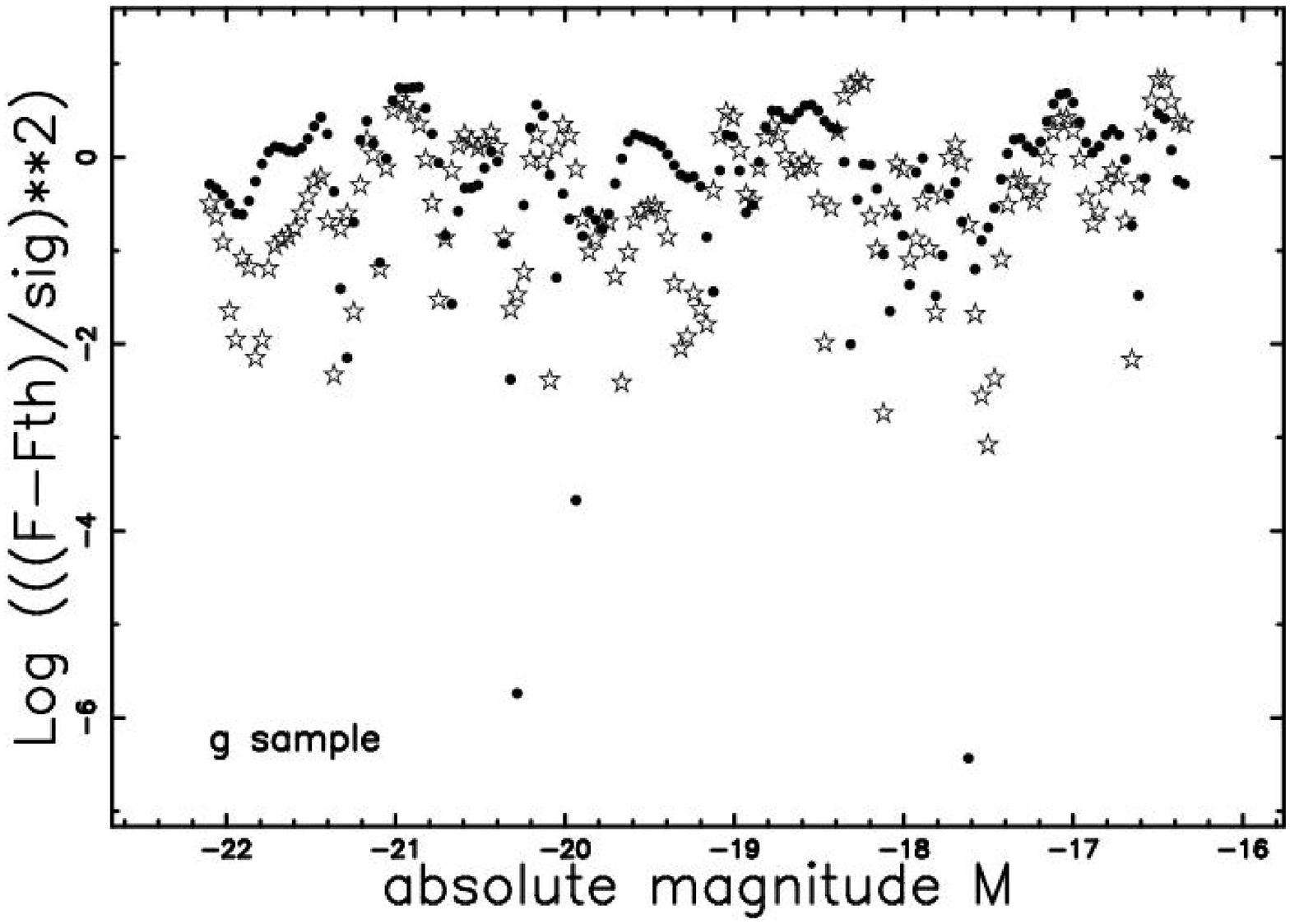}
\caption {The residuals of  the fits to 
SDSS($g^*$) data.
 The empty stars represent  our  
two luminosity functions 
((\ref{equation_mia}) and (\ref{equation_miapareto}) )  
and the filled points 
represent   the Schechter function.
 }
          \label{residui_g} 
    \end{figure}

\begin{figure}
   \centering
\plotone{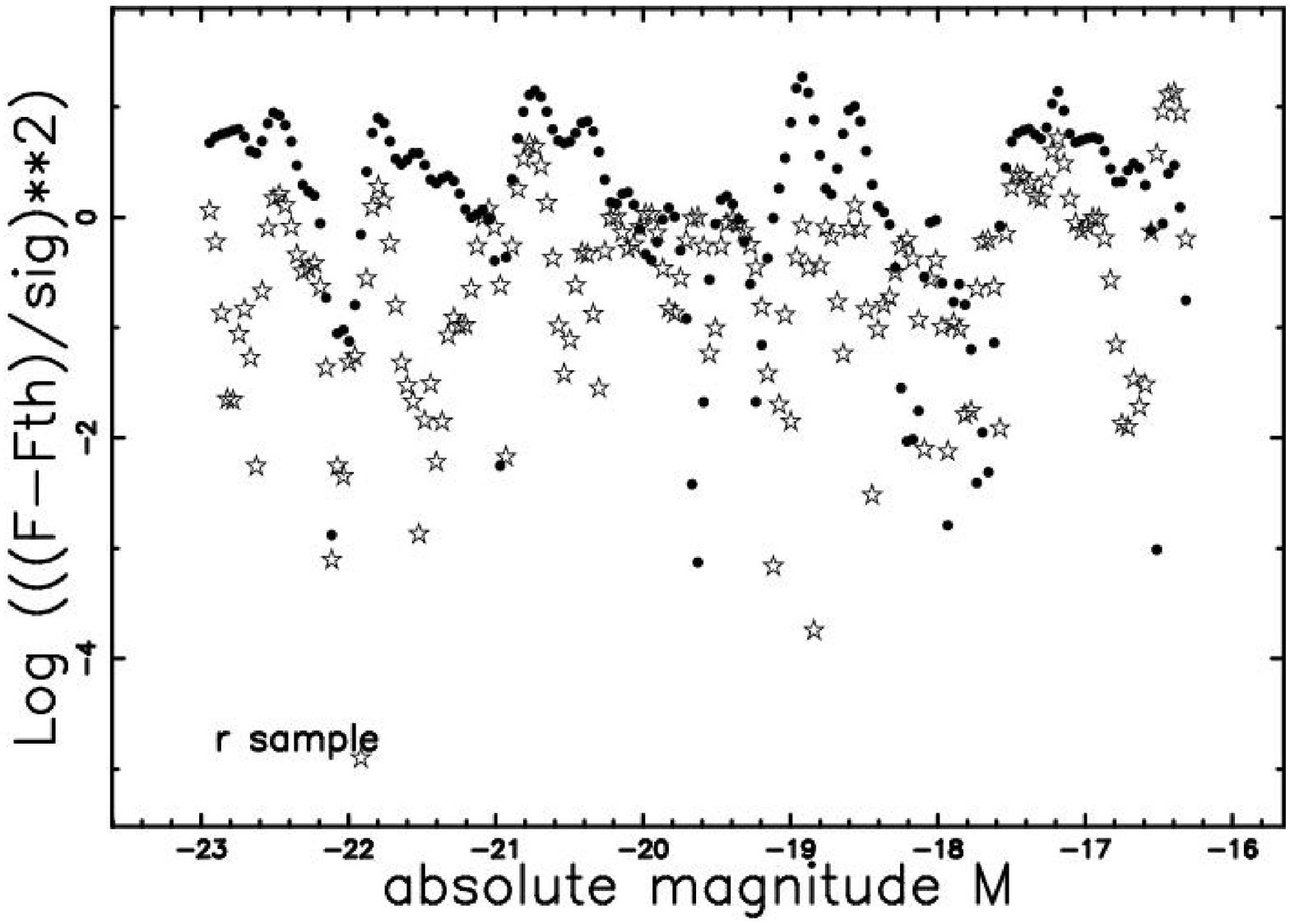}
\caption {The residuals of  the fits to 
SDSS($r^*$) data.
 The empty stars represent  our  
two luminosity functions 
((\ref{equation_mia}) and (\ref{equation_miapareto}) )  
and the filled points 
represent   the Schechter function.
 }
          \label{residui_r} 
    \end{figure}

\begin{figure}
   \centering
\plotone{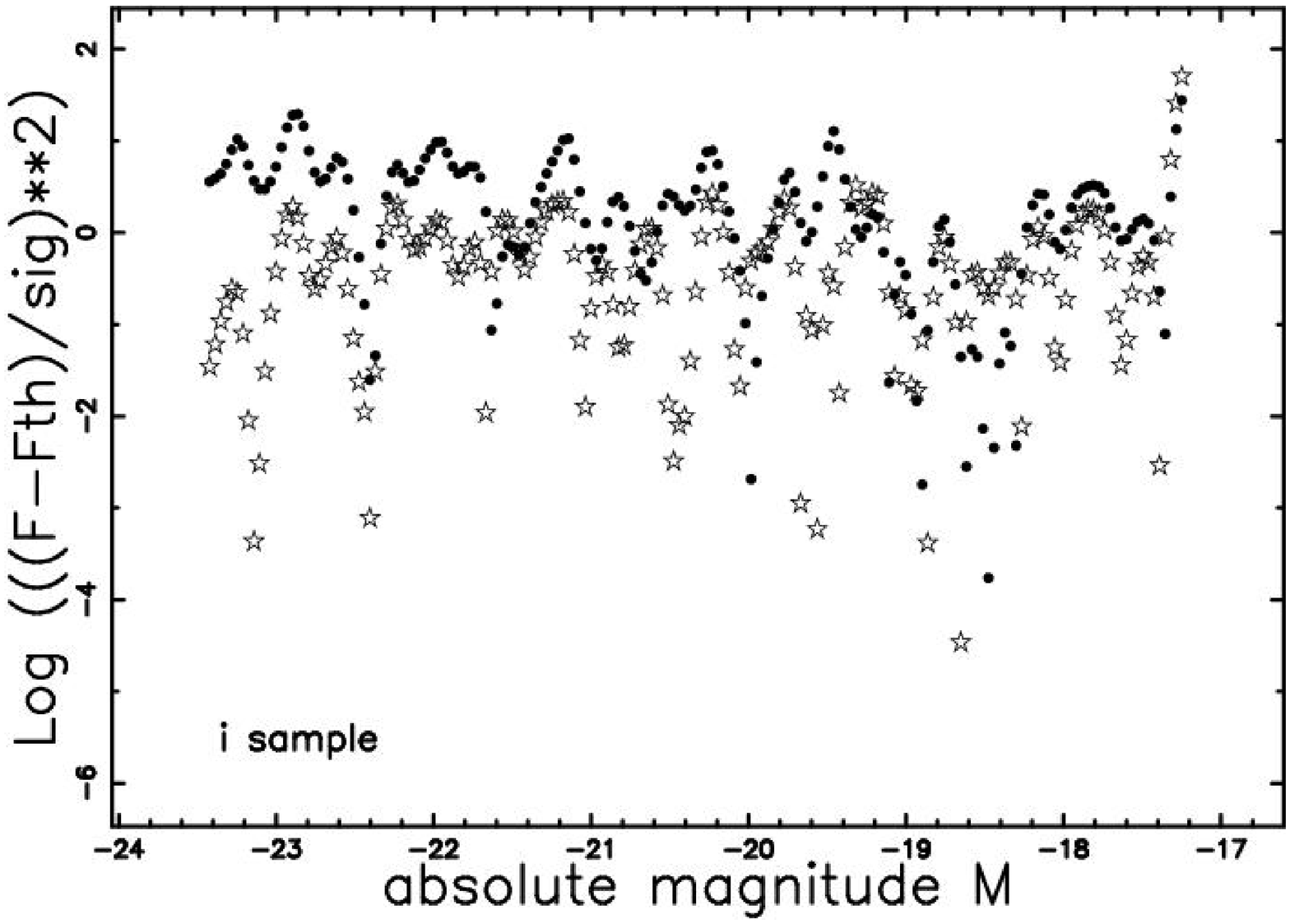}
\caption {The residuals of  the fits to 
SDSS($i^*$) data.
 The empty stars represent  our  
two luminosity functions 
((\ref{equation_mia}) and (\ref{equation_miapareto}) )  
and the filled points 
represent   the Schechter function.
 }
          \label{residui_i} 
    \end{figure}

\begin{figure}
   \centering
\plotone{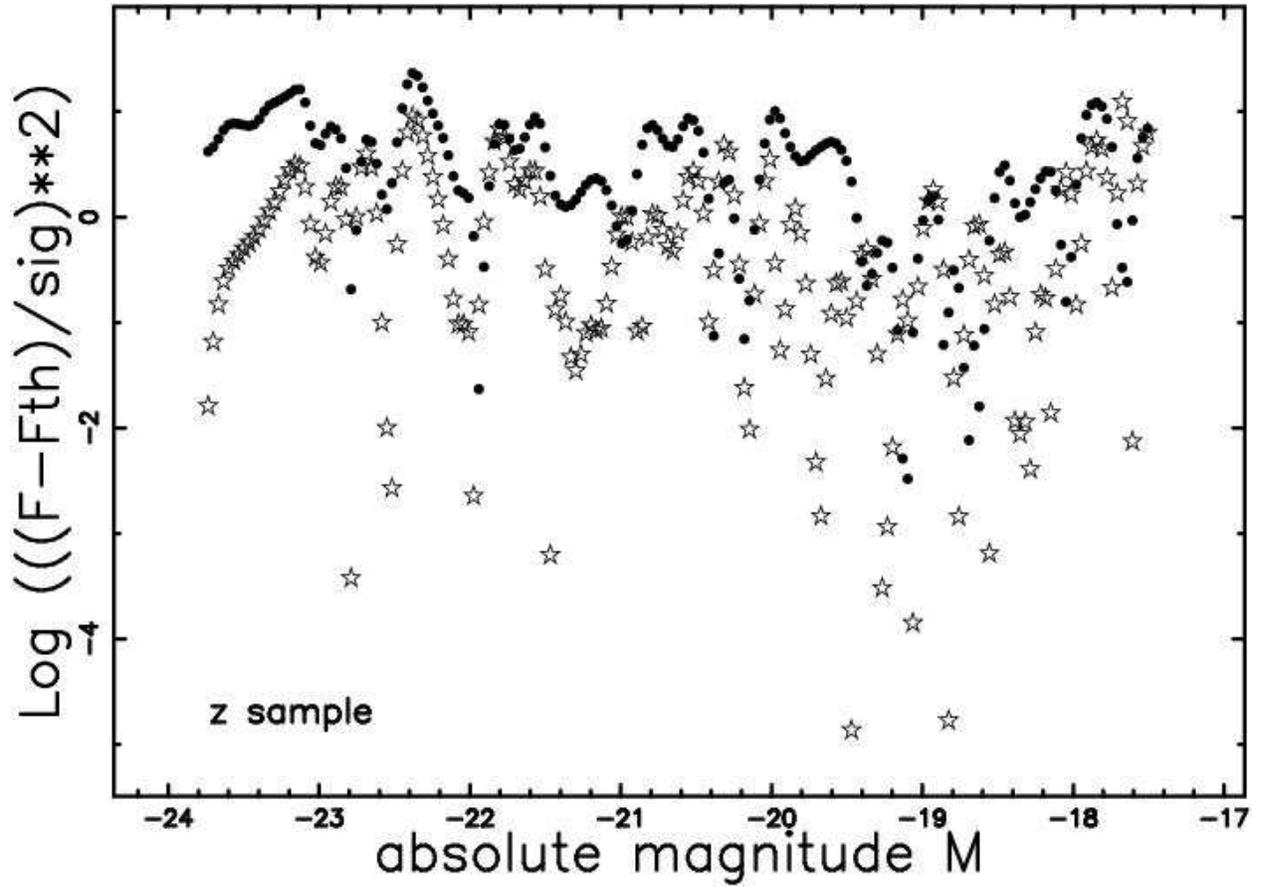}
\caption {The residuals of  the fits to 
SDSS($z^*$) data.
 The empty stars represent  our  
two luminosity functions 
((\ref{equation_mia}) and (\ref{equation_miapareto}) )  
and the filled points 
represent   the Schechter function.
 }
          \label{residui_z} 
    \end{figure}

The Schechter function , the new  function and the data are
reported in 
Figure~\ref{due_u} ,
Figure~\ref{due_g} ,
Figure~\ref{due_r} ,
Figure~\ref{due_i} ,
and Figure~\ref{due_z}
 when the $u^*$,$g^*$ ,$r^*$ , $i^*$  and $z^*$ 
 bands  are  considered;
    Figure~\ref{residui_u} ,
    Figure~\ref{residui_g} ,
    Figure~\ref{residui_r} ,
    Figure~\ref{residui_i} ,
and Figure~\ref{residui_z}
 report the residuals  of the 
 $u^*$,$g^*$ ,$r^*$ , $i^*$  and $z^*$ band.
We have used  $H_0 = 100 h$ km s$^{-1}$ Mpc$^{-1}$, with $h=1$
in all the numerical evaluations.
Due to the testing phase of the new $\mathcal {M}-L$  function,
we have omitted the propagation of 
other values of  $h$  on the derived quantities ,
see the discussion in~\citet{blanton}.

\clearpage

\clearpage

\clearpage

\clearpage

\clearpage

\clearpage

\clearpage

\clearpage

\clearpage

\clearpage

\clearpage

The value 
obtained for the parameter {\it a }  should be   
compared with that of the normal stars  which   is 3 or  4 as suggested
by  the theory, see for example
\citet{lang}, 
or    $\approx$ 3.8 as  suggested by the observations
for ${\mathcal M}~>~0.2{\mathcal M}_{\sun}$, see
for example \citet{cox}. 
When  the three 
classes  of stars are considered we have
{\it a} =3.43 (MAIN),{\it a} =2.79 (GIANTS) and  {\it a}=2.43 (SUPERGIANTS),
see~\citet{zaninetti05} for further details.

The variation of $j$ when the range in magnitude is finite
rather than infinite can be evaluated by  coupling 
together  formula~(\ref{mlrelation})  and 
(\ref{equation_mia})
\begin{equation}
j = \int_{-21}^{-15.78}  10^{0.4(M_{bol,\sun} - M)}
\Psi (M) dM 
\quad .
\label{jfinite}
\end{equation}
On inserting the parameters of  SDSS band $u^*$ (
which  is the case in which the $\mathcal {M}-L$  function   
covers all the range in magnitude of the data 
, see Table~\ref{para_physical_SDSS_data} )
and  ${M}_{bol,\sun}$=${M}_{u^*\sun}$=6.39
, $j=1.4\;10^{8}\; L_{\sun} $ is obtained.
This value increases by  $5.63\;\%$  when the range is infinite;
see a similar discussion concerning 
Schechter's function  around formula~(33) in \citet{lin}.

In   absence of observational  data 
that represent  the luminosity function,
 we can generate them 
through  Schechter's parameters, see Table~\ref{parameters}.
This is done,  for example for 
the 2dF Galaxy
Redshift Survey (2dFGRS) ,see~\cite{Cross2001}
and data of  the  Schechter function in  Table~\ref{parameters}. 
The parameters  of the ${\mathcal M}-L$ function
are reported in Table~\ref{para_physical}
where  the requested errors on the
values of luminosity are   the same as    the considered value.
 \begin{table} 
 \caption[]{The parameters of the ${\mathcal M}-L$ luminosity function \\
            based on  2dFGRS data 
           ( triplets  generated by the author) } 
 \label{para_physical} 
 \[ 
 \begin{array}{lc} 
 \hline 
~     &   2dFGRS   \\ \noalign{\smallskip}  
 \hline 
 \noalign{\smallskip} 
c                   &    0.1                 \\ \noalign{\smallskip}
M^*    [mags]       &  -19  \pm 0.1       \\ \noalign{\smallskip}
\Psi^* [h~Mpc^{-3}] &  0.4  \pm 0.01      \\ \noalign{\smallskip}
a                   &  1.3   \pm 0.1       \\ \noalign{\smallskip} 
 \hline 
 \hline 
 \end{array} 
 \] 
 \end {table}
\section{Tests involving z}
\label{secz}
Some useful formulae connected with the 
Schechter function in a Euclidean ,non-relativistic 
and homogeneous universe 
are reviewed ;
by analogy new formulae for  
the first part of the 
${\mathcal M}-L $  function are derived.

\subsection{The behaviour of the Schechter function}

The flux of radiation , {\it f} , is  introduced 
\begin{equation}
f  = \frac{L}{4 \pi r^2} 
\quad ,
\end{equation}
where {\it r}  represents the distance of the galaxy .
The joint distribution in {\it z}  and {\it f}  for galaxies ,
see formula~(1.104) in~\cite{pad} ,
 is
\begin{equation}
\frac{dN}{d\Omega dz df} =  4 \pi 
\bigl ( \frac {c}{H_0} \bigr )^5    z^4 \Phi (\frac{z^2}{z_{crit}^2})
\label{nfunctionz}  
\quad ,
\end {equation}
where $d\Omega$ , $dz$ and  $ df $ represent the differential of
the solid angle , the red-shift and the flux respectively.
The $L^*$ of difference between the previous formula 
and formula~(1.104) in~\cite{pad} is due to the small difference
in the definition of $\Phi$.

\begin{figure}
\plotone{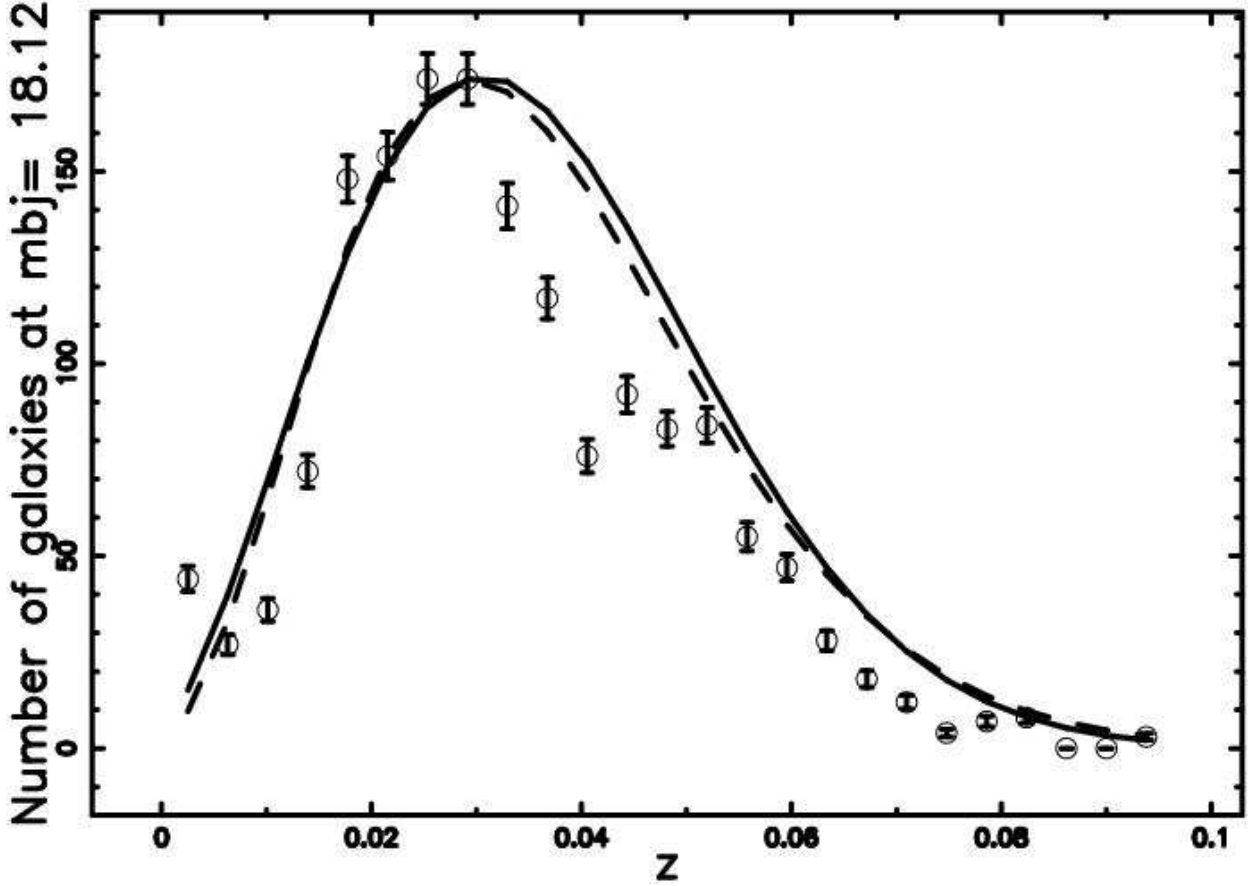} 
\caption{
The galaxies  of the 2dF Galaxy Redshift Survey with 
$ 15.08  \leq  bJmag \leq 15.81 $
( with $bJmag$ representing  the 
relative magnitude  used in object selection),
are isolated 
in order to represent a chosen value of $f$ 
and then organised in frequencies versus
heliocentric  redshift
 ,  (empty circles);
the error bar is given by the square root of the frequency.
The theoretical curve  generated by
the Schechter  function of luminosity 
(formula~(\ref{nfunctionz}) and parameters
as in column 2dFGRS of Table~\ref{parameters}) 
is drawn  (full line).
The theoretical curve  generated by
the ${\mathcal M}-L$   function for luminosity (
formula~(\ref{nfunctionz_mia})
and  parameters as in column  2dFGRS of Table~\ref{para_physical})
is drawn  (dashed line);
 $\chi^2$= 320  for the Schechter  function and $\chi^2$= 283
for the ${\mathcal M}-L$   function.
}
          \label{maximum}%
    \end{figure}

The formula  for   $z_{crit}$ is 
\begin{equation}
 z_{crit}^2 = \frac {H_0^2  L^* } {4 \pi f c_L^2}
\quad ,
\end{equation} 
where $c_L$ represents the light velocity;
the CODATA recommends  $c_L  = 299792.458 \frac{km}{s} $ .
The mean red-shift of galaxies with a flux  $f$ 
,
see formula~(1.105) in~\cite{pad} ,
 is
\begin{equation}
\langle z \rangle = z_{crit}  \frac {\Gamma (3 +\alpha)} {\Gamma (5/2 +\alpha)}
\quad .
\end{equation} 
The number density of galaxies per unit flux interval,
see formula~(1.106) in~\cite{pad} , is
\begin{equation}
\frac {dN}{d ln f } = \frac {\Phi^*}{2}  
\bigl ( \frac {L^*}{4\pi f} \bigr )^{3/2} \Gamma (\frac{5}{2} + \alpha)  
\quad  .
\end{equation}
The number of galaxies in {\it z} and {\it f} as given by 
formula~(\ref{nfunctionz})  has a maximum  at  $z=z_{max}$ ,
where 
\begin{equation}
 z_{max} = z_{crit}  \sqrt {\alpha +2 }
\quad .
\end{equation} 
The value of  $z_{max}$ can be derived from the 
 histogram of the observed number of galaxies expressed as a function
of $z$ .
For  practical purposes we analysed  
the 2dFGRS data release 
available at  the  web site: http://msowww.anu.edu.au/2dFGRS/.
In particular we added together the file parent.ngp.txt that 
contains 145652 entries for NGP strip sources and 
the file parent.sgp.txt that 
contains 204490 entries for SGP strip sources.
Once the   heliocentric red-shift  was  selected 
we  processed 219107 galaxies with 
$0.001 \leq z \leq 0.25$ .
A comparison between the observed and theoretical number of galaxies
as  a function of $z$ is reported in Figure~\ref{maximum}.

\clearpage

\begin{figure}
\plotone{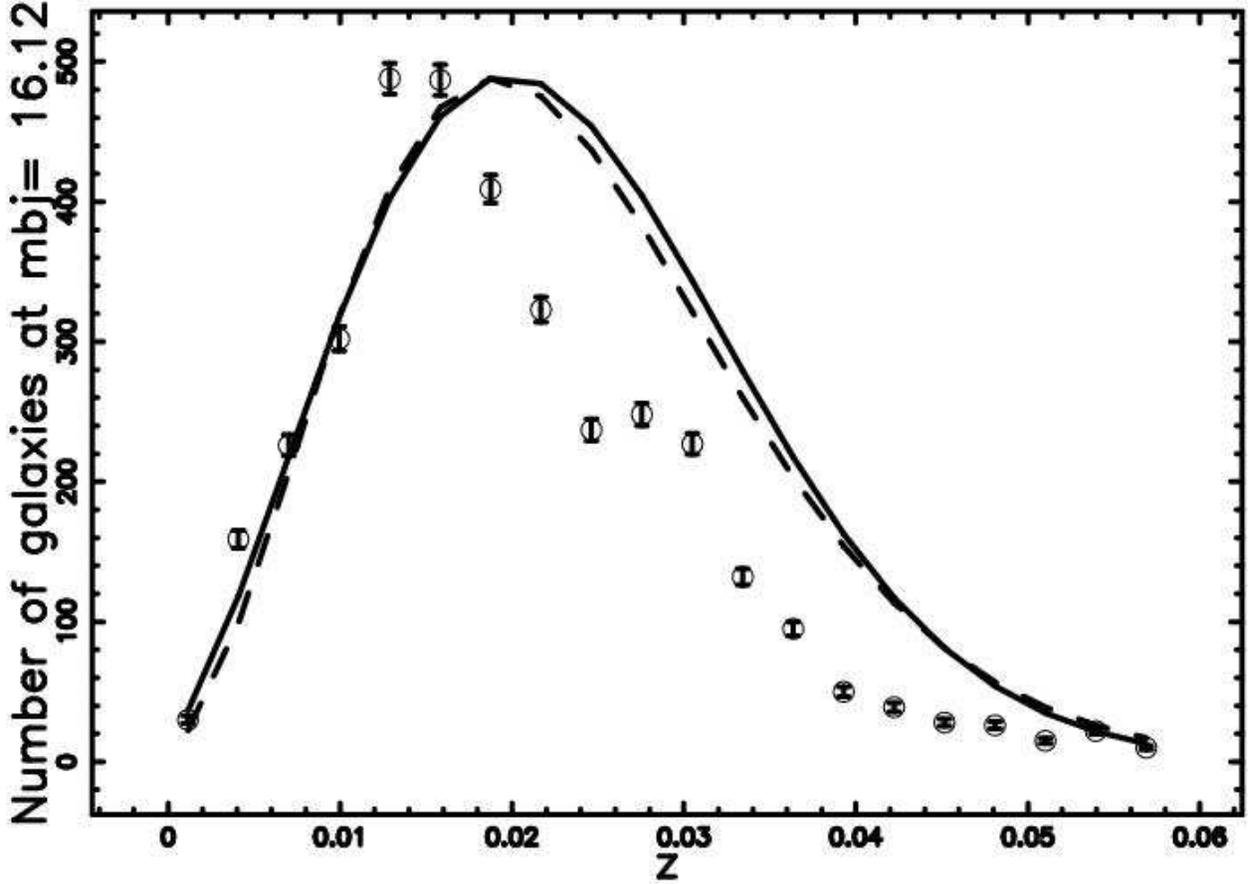} 
\caption{The galaxies  of the 6dF Galaxy  Survey with 
$ 14.15  \leq  bJmag \leq 14.9 $
( with $bJmag$ representing  the 
relative magnitude  used in object selection),
are isolated 
in order to represent a chosen value of $f$ 
and then organised in frequencies versus
 redshift
 ,  (empty circles);
the error bar is given by the square root of the frequency.
The theoretical curve  generated by
the Schechter  function of luminosity 
(formula~(\ref{nfunctionz}) and parameters
as in column 2dFGRS of Table~\ref{parameters}) 
is drawn  (full line).
The theoretical curve  generated by
the ${\mathcal M}-L$   function for luminosity (
formula~(\ref{nfunctionz_mia})
and  parameters in column  2dFGRS of Table~\ref{para_physical})
is drawn  (dashed line);
 $\chi^2$= 1373 for the Schechter  function and $\chi^2$= 1197 
for the ${\mathcal M}-L$   function.
}
          \label{maximum_6d}%
    \end{figure}

Another interesting catalog  is the 
 6dF Galaxy Survey that has  measured 
  around 150000 redshifts and 15000
    peculiar velocities from galaxies over the southern sky ,
see~\cite{Jones2006}.
It is available  at the following address 
http://vizier.u-strasbg.fr/viz-bin/VizieR?-source=VII/249
 ;  we selected the re-calibrated $bJ$ magnitude and 
the recession velocity cz.
Figure~\ref{maximum_6d} reports  the observed 
and theoretical number of galaxies
as  a function of $z$ for the  6dF Galaxy Survey.

\clearpage

\subsection{The behaviour of the ${\mathcal M}-L$  function}

The joint distribution in $z$ and $f$ ,
in presence of the ${\mathcal M}-L$
luminosity (equation~(\ref{equation_schechter_mia})) is 
\begin{equation}
\frac{dN}{d\Omega dz df} =  4 \pi 
\bigl ( \frac {c}{H_0} \bigr )^5    z^4 \Psi (\frac{z^2}{z_{crit}^2})
\label{nfunctionz_mia}  
\quad .
\end {equation}
The mean red-shift  is  
\begin{equation}
\langle z \rangle = z_{crit} 
\frac {2\;\;{4}^{-{\frac {2\,a+{\it c}}{a}}}\Gamma
\left( 2\,a+{\it c}
 \right) {2}^{{\frac {2\,{\it c}+3\,a}{a}}} }{ \Gamma  \left( {
\it c}+3/2\,a \right)   }
\quad .
\label{zmediomia}
\end{equation} 
The number density of galaxies per unit flux interval is
\begin{equation}
\frac {dN}{d ln f } =
\frac{1}{16} \,{\frac {{{\it L_*}}^{3/2}{\it \Psi_*}\,\Gamma  \left( {\it c
}+\frac{3}{2}\,a \right) }{{\pi }^{3/2}{f}^{3/2}\Gamma  \left( {\it c}
 \right) }}
\quad  .
\end{equation}
The number of galaxies as  given 
by formula~(\ref{nfunctionz_mia}) has a maximum at 
$z_{max}$  where 
\begin{equation}
 z_{max} = z_{crit} 
\left( {\it c}+a \right) ^{a/2}
\quad .
\end{equation} 
A comparison between the observed and theoretical number of galaxies
as given by  the ${\mathcal M}-L$ function 
is reported in Figure~\ref{maximum} where  the 
2dF Galaxy Redshift Survey  is considered 
and in Figure~\ref{maximum_6d} where  the 
6dF Galaxy  Survey   is considered.

\section{Mass evaluation}
\label{secmass}

One method to deduce the mass of a star by its absolute visual
magnitude is presented; the mass   of a galaxy  is deduced by
analogy. In the case of the galaxies,
 the bolometric correction  of
the stars will be replaced by  the sun's absolute  magnitude   and
mass-luminosity ratio different  in each  selected band.

\subsection{The case of the stars}

In the case of the stars it is possible to parameterise
the mass of the star , ${\mathcal M_S}$ ,
as a function of the observable colour $(B-V)$
, see~\cite{zaninetti05} ~.
The first equation connects
 the  $(B-V)$ colour  with the temperature
\begin{equation}
(B-V)=  K_{\mathrm{BV}}   + \frac {T_{\mathrm{BV}}}{T}
\quad,
\label{bvt}
\end {equation}
here $T$ is  the temperature, $K_{\mathrm{BV}}$  and $T_{\mathrm{BV}}$
are two parameters that  can be derived by implementing
the least square method  on a series of calibrated   data.
 The  second equation describes the bolometric
correction , $BC$,
\begin {equation}
BC  = M_{\mathrm{bol}}  -M_{\mathrm{V}} =
-\frac{T_{\mathrm{BC}}}{T} - 10~\log_{10}~T + K_{\mathrm {BC}}
\quad,
\label{bct}
\end {equation}
where $M_{\mathrm{bol}}$   is the absolute bolometric magnitude,
      $M_{V}$     is the absolute visual     magnitude,
      $T_{\mathrm {BC}}$ and  $K_{\mathrm{BC}}$ are two parameters that can be derived
      through the general linear least square  method
      applied to a series of calibrated data.
The third equation is
 the usual formula for the luminosity
\begin{equation}
\log_{10}(\frac {L}{L_{\sun}})  =  0.4 (4.74 -M_{\mathrm{bol}})
\quad,
\end {equation}
where $L$ is the luminosity of the star and
${L_{\sun}}$ the luminosity of the sun.
The fourth equation is the usual mass-luminosity
 relationship for stars
\begin{eqnarray}
\log_{10}(\frac {L}{L_{\sun}})  = a_{\mathrm {LM}} +b_{\mathrm {LM}}\log_{10}(\frac {{\mathcal
M}_S}
 {{\mathcal M}_{\sun}}) \\
 for~{\mathcal M} > 0.2 {\mathcal {M}}_{\sun} \nonumber
\quad,
\end {eqnarray}
where $\mathcal{M}_S$  is the  mass of the star and
$\mathcal{M}_{\sun}$  is the  mass of the sun.

With these  four equations   the mass of the star is
\begin{eqnarray}
  \log_{10} \frac {{\mathcal M}_S}
 {{\mathcal M}_{\sun}}=   \nonumber
 \end{eqnarray}
 \begin{eqnarray}
\frac {- 0.4\,{\it M_{\mathrm V}}- 0.4\,{\it K_{\mathrm {BC}}}+ 4.0\,
 \ln
 \left( {\frac {{\it T_{\mathrm{BV}}}}{{\it (B-V)}-{\it K_{\mathrm{BV}}}}} \right)  \left( \ln
 \left( 10 \right)   ^{-1}
  \right )}{b_{\mathrm {LM}}}  \nonumber
  \end{eqnarray}
 \begin{eqnarray}
 -\frac{
  \ 0.4\,{\frac {{\it T_{\mathrm {BC}}}\,
 \left( {\it -(B-V)}+{\it K_{\mathrm{BV}}} \right)}{{\it T_{\mathrm{BV}}}}}+
1.896-{\it a_{\mathrm {LM}}
}} {b_{\mathrm {LM}}}
\label{mass_analytical}
\quad,
\end {eqnarray}
with the  various coefficients  as given by  Table~1
in~\cite{zaninetti05}.
As an example, the mass of a star belonging
to MAIN SEQUENCE V  is
\begin{equation}
  \log_{10} \frac {{\mathcal M_S}}
 {{\mathcal M}_{\sun}}=
- 7.769+ 0.8972\,\ln  \left(  \frac{ 7361} {{\it (B-V)}+
 0.6411 } \right)
\quad .
\end{equation}

We can now express the colour $(B-V)$ as a function of the
absolute visual     magnitude $M_{V}$
and the following formula for the mass of the star is obtained
\begin{eqnarray}
  \log_{10} \frac {{\mathcal M_S}}
 {{\mathcal M}_{\sun}}=
- 7.769+ 0.8972\,\ln  \left( \frac { 9378} {  {\it
W} \left(  9378\,{e^{- 8.496+ 0.2972\,{\it
M_V}}}   \right) } \right)
\\
MAIN~SEQUENCE~V~ when   ~-5.8 < M_V < 11.8
\quad ,  \nonumber
\end{eqnarray}
where $W$ is the Lambert~W-function, after~\cite{Lambert_1758}. A test
of the previous formula can be done at the two boundaries: when
$M_V$ =-0.58 , $ \log_{10} \frac {{\mathcal M_S}}{{\mathcal
M}_{\sun}}=1.63$ against the calibrated value $\log_{10} \frac
{{\mathcal M_S}}{{\mathcal M}_{\sun}}=1.6$ and when $M_V$ =11.8 ,
$ \log_{10} \frac {{\mathcal M_S}}{{\mathcal M}_{\sun}}=-0.56$
against the calibrated value $\log_{10} \frac {{\mathcal
M_S}}{{\mathcal M}_{\sun}}=-0.66$, see Table~3.1
in~\cite{deeming}.

\subsection{The case of the galaxies}

\begin{figure}
\begin{center}
\plotone{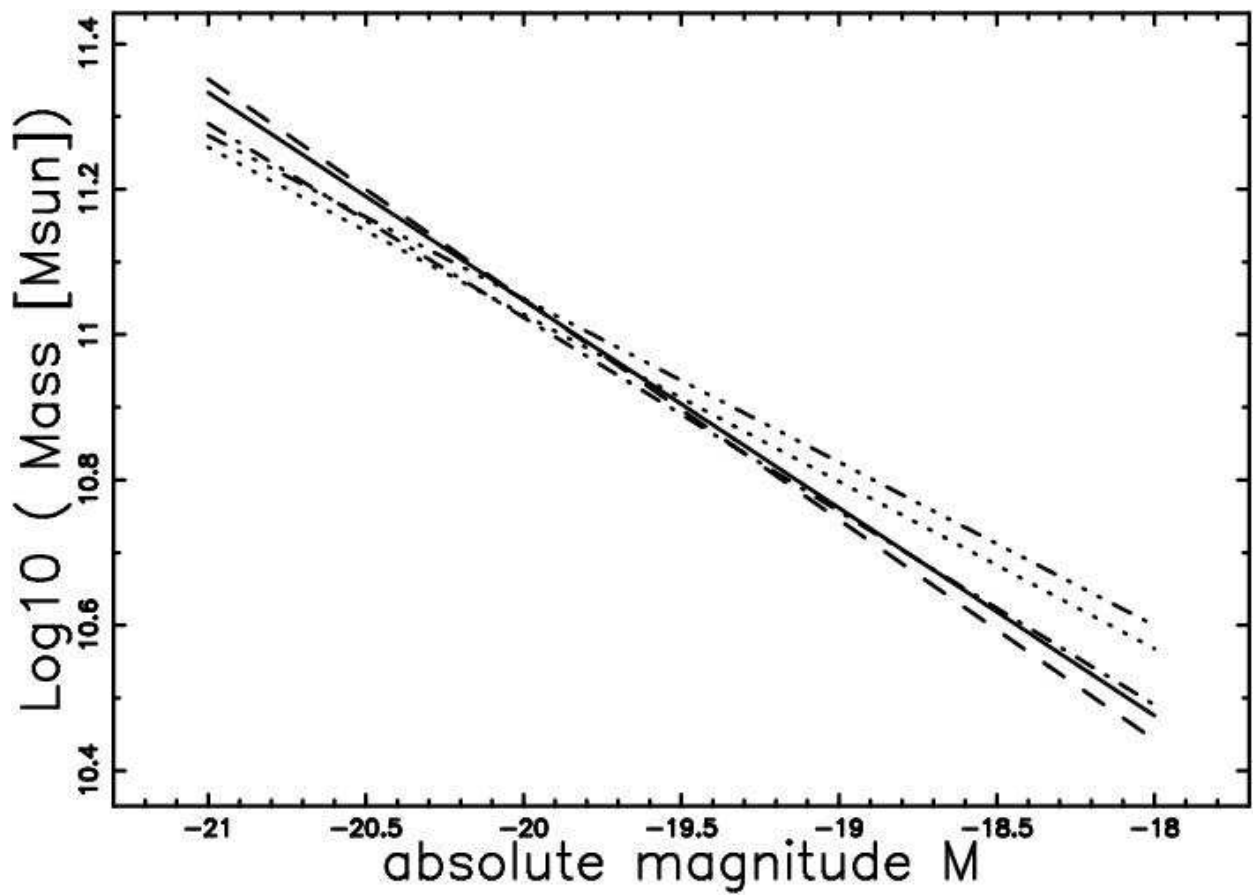} 
\end {center}
\caption { 
Logarithm of the mass of the galaxy as a function of the absolute magnitude.
The SDSS bands are  
$u^*$ with ${M}_{bol,\sun}$=${M}_{u^*\sun}$=6.39 
and $R=6$ (full line) , 
$g^*$ with ${M}_{bol,\sun}$=${M}_{g^*\sun}$=5.07
and $R=13$  (dashed), 
$r^*$ with ${M}_{bol,\sun}$=${M}_{r^*\sun}$=4.62
and $R=16$  (dot-dash-dot-dash), 
$i^*$ with ${M}_{bol,\sun}$=${M}_{i^*\sun}$=4.52 
and $R=15$ (dotted),
$z^*$ with ${M}_{bol,\sun}$=${M}_{z^*\sun}$=4.48 
and $R=14$ 
(dash-dot-dot-dot). 
}
          \label{masse}%
    \end{figure}

The mass of a galaxy can be evaluated once the mass luminosity
ratio , $R$ is given
\begin{equation}
R = \langle   \frac{M}{L} \rangle
\quad .
\end{equation}
Some values of  $R$ are now reported :
$ R \leq 20$ by~\cite{kiang1961} and \cite{Persic_1992} ,
$R =20$ by~\cite{pad} , 
$R =5.93 $ by~\cite{vandermarel1991}.
Further on \cite{Bell2001},  demonstrated (amongst others) that
$\frac{\mathcal{M}}{L} $ 
varies as a function of galaxy colour, and therefore, type.
If the bright end of the luminosity function is dominated 
by massive, evolved, red galaxies,
and the faint end by low mass,  blue galaxies, then
$\frac{\mathcal{M}}{L}\propto L^{-0.64}$ (GIANTS III) 
at the bright end  and 
$\frac{\mathcal{M}}{L}\propto L^{-0.7}$ (MAIN SEQUENCE V) 
at the faint end , see coefficients of  Table~1 in~\cite{zaninetti05}.
Then $\frac{\mathcal{M}}{L} $ 
will almost certainly not be constant  due to different prevailing
populations  of stars at the boundaries  of the luminosity function.
The scatter in the models  by \cite{Bell2001} 
is a starting point when evaluating the validity of
assuming a constant $\frac{\mathcal{M}}{L} $. 
Generally, near-infrared $\frac{\mathcal{M}}{L} $ ratios
are more constant than   optical passband, but still vary with luminosity.
In our framework  , 
we  made $R$  a  function of the passband ,
in order to have   the same  results for the 
masses  of the galaxies  once  the  absolute magnitude  is given, 
 see Figure~\ref{masse}.
In our framework $R$ can be expressed as
\begin{equation}
R = \frac { \langle{\mathcal M}\rangle }{ \langle L \rangle } 
\quad .
\end{equation}
On inserting formula~(\ref{massamedia}) 
and  formula (\ref{lmedia}) in the previous ratio   
the following formula  for  ${\mathcal M^*}$ is found 
\begin{equation}
{\mathcal M}^* = R L^* \frac{\Gamma (c+a)}{\Gamma(c)}
\frac { {\mathcal M}_{\sun}} {L_{\sun}} 
\label{mstar}
\quad .
\end {equation}
From equations~(\ref{massa})  and~(\ref{mstar})
a formula  for the mass of the galaxy is found
\begin{equation}
\mathcal{M} = 
{\frac {R{10}^{( 0.4\,{\it M_{bol,\sun}}- 0.4\,{\it M^*})}\Gamma 
 \left( c+a \right)  \left( {10}^{- 0.4\,{\it M}+ 0.4\,{\it M^*}} \right) ^{{a}^{-1}}}{c\Gamma  \left( c \right) }}
{ {\mathcal M}_{\sun}}
\quad .
\label{massgalaxy}
\end{equation}

An  application of the previous 
formula is reported in Figure~(\ref{masse}) ,
where  the mass of galaxies as a function of the absolute
magnitude in the five bands of SDSS is drawn.
In this Figure ${M}_{bol,\sun}$ is  different 
for each selected band and equal to the value
suggested in equation~(16) of~\cite{blanton}.

\clearpage

\clearpage

      The  new  formula~(\ref{massgalaxy})  allows us to deduce 
      the mass of the galaxy from its absolute magnitude
      and can be easily particularized in different pass-bands.
      As an  example , with the data of  SDSS in the
      five  bands  reported in Table~\ref{para_physical_SDSS_data}
      ,  ${M}_{bol,\sun}$ as 
      in equation~(16) of~\cite{blanton} 
      and  $R$ as in Figure~\ref{masse},
we have
\begin{eqnarray}
\mathcal{M}=&
 215234\,{e^{- 0.6579\,{\it M}}} 
   {{\mathcal M}_{\sun}}  & ~u^*~band
~when  -20.6 \leq M \leq  -15.7  
\nonumber \\
\mathcal{M}=&
97216\,{e^{- 0.6978\,{\it M}}}
   {{\mathcal M}_{\sun}}  & ~g^*~band
~when  -22.0  \leq M \leq   -18.2
 \nonumber \\
\mathcal{M}=&
490000\,{e^{- 0.6141\,{\it M}}}
   {{\mathcal M}_{\sun}}  & ~r^*~band 
~when  -23.0  \leq M \leq   -19 
\label{massparticular}
         \\
\mathcal{M}=&
2691000\,{e^{- 0.5294\,{\it M}}}
   {{\mathcal M}_{\sun}}  & ~i^*~band 
~when  -23.5  \leq M \leq   -19.3  
\nonumber \\
\mathcal{M}=&
3434000\,{e^{- 0.5175\,{\it M}}}
   {{\mathcal M}_{\sun}}  & ~z^*~band 
~when  -23.7  \leq M \leq   -20.0
\nonumber \quad .
\end{eqnarray}
      The method here suggested to deduce the mass of the galaxy
      can be compared with the formula  that comes out 
      from the Tully-Fisher relation , see \cite{Tully1977} 
      and  \cite{Tully1998}.
      In the  Tully-Fisher  framework the mass of a rotating 
      galaxy  can be parameterised 
      as  
\begin{equation}
\mathcal{M}= 50  {V_f}^4 { {\mathcal M}_{\sun}}   \frac{s^4}{Km^4} 
\quad ,
\label{eqnfisher}
\end{equation}
 where $V_f$ is the rotational velocity expressed in
 $\frac{Km}{s}$, see~\cite{McGaugh2005}. 
        The mass to light ratio in our framework scales
        $\propto L^{(1/a) -1}$ with $a$
        depending  on  the selected catalog and band.
        This  ratio oscillates , referring to the SDSS data,
        between  a minimum 
        dependence in the $g^*$ band , 
        $\frac{\mathcal{M}}{L}\propto L^{-0.24}$
        and a maximum dependence in the 
        $i^*$ band , $\frac{\mathcal{M}}{L}\propto L^{-0.42}$.
        A comparison should be made  with 
        $\frac{\mathcal{M}}{L}\propto L^{0.35}$
        in \cite{vandermarel1991} for a sample of 
        37  bright elliptical  galaxies ; this result was
        obtained by implementing axisymmetric dynamical models. 
The completeness of the mass sample of the galaxies belonging to a given catalog
 can be evaluated 
in the following way.
The limiting apparent magnitude 
is known, 
and is different for each catalog.
In the case of the SDSS ($r^*$ band) 
is  $m$=17.6, see~\cite{blanton}.

The corresponding absolute limiting magnitude is computed 
and inserted in equation~(\ref{massgalaxy}).  
The limiting 
mass for galaxies , ${\mathcal M}_L$  ,is
\begin{equation}
{\mathcal M}_L =
R \frac{{10}^{( 0.4\,{\it M_{bol,\sun}}- 0.4\,{\it M^*}\/)}\Gamma  \left( c+a
 \right)  \left({10}^{- 0.4\,{\it m }+ 2.0\,{\it Log10}
 \left({\frac{{\it c_L z}}{H_{{0}}}} \right) + 10.0+ 0.4\,{\it 
M^*}} \right) ^{{a}^{-1}} }
{{c} \left( \Gamma  \left( c
 \right)  \right) }
{\mathcal M_{\sun}} 
\quad ,
\label{limitingmass}
\end{equation}
where  $c_Lz$ is the radial distance expressed in  
$km/s$.
In order to  see how the parameter $z$ influences  
the limiting mass, 
Table~\ref{para_limiting} reports  the range of observable 
masses as a function of $z$.

\begin{table} 
\caption [] {The limiting mass for the SDSS catalog ,$u^*$ band,
when $R$  and   $M_{bol,\sun}$  
are those of Figure~\ref{masse}.
The limiting apparent magnitude is  $m$=17.6.}
 \label{para_limiting} 
 \[ 
 \begin{array}{lc} 
 \hline 
mass~range     &   z  \\ \noalign{\smallskip}  
 \hline 
 \noalign{\smallskip} 
  1.41 10^8{\mathcal M}_{\sun} <{\mathcal M}    < 1.7~10^{11}{\mathcal M}_{\sun}  &    
  0.001  \\ \noalign{\smallskip}
  3.79 10^{9}{\mathcal M}_{\sun} <{\mathcal M}  <  1.7~10^{11}{\mathcal M}_{\sun}  &
  0.01  \\ \noalign{\smallskip}
  1.0  10^{11}{\mathcal M}_{\sun} <{\mathcal M}  < 1.7~10^{11}{\mathcal M}_{\sun}  &    
  0.1  \\ \noalign{\smallskip}
  1.4 10^{11}{\mathcal M}_{\sun} <{\mathcal M}  < 1.7~10^{11}{\mathcal M}_{\sun}  &    
  0.13  \\ \noalign{\smallskip}
 \hline 
 \hline 
 \end{array} 
 \] 
 \end{table} 

\section{Conclusions}

\label{conclusions}
We have split the analysis of the  luminosity function in
two .
The analysis of the main new luminosity function , 
formula~(\ref{equation_mia}),
 from 
low luminosities up  to  maximum magnitude
shows that , see   Table~\ref{para_physical_SDSS_data},
\begin{enumerate} 
\item   The parameter {\it c } varies between 0.1  and 2.
        It must be remembered  that the theory predicts 
        2 , 4 and 6 for
        the 1D,2D and 3D fragmentation  respectively.
\item   Parameter  {\it  a } varies between 1.32  and 1.74.
        The numerical mass-luminosity relationship 
        for the stars  gives values of the parameter {\it a}
        comprised between 2.43 and 3.43.
\item   The  ${\mathcal M} -L$  function
        represents a better fit of the observational data 
        in comparison with  the Schechter function  once the concept
        of  maximum magnitude of the sample is introduced.
        Without this limiting magnitude the situation is inverted.
\end  {enumerate}

The case of low luminosities galaxies was 
describe by a truncated Pareto type luminosity
function , see formula~(\ref{equation_miapareto}).
This new luminosity function is described by two 
physical parameters , $d$ , and $a$ denoting 
respectively the distribution in mass and 
the mass-luminosity connection.
The analysis of the data 
for low luminosities galaxies
as reported in Table~\ref{para_physical_SDSS_LL}
shows that 
\begin{enumerate} 
\item  The parameter  $d$ varies between 0.3 and 0.9~.
       This value should be compared with 
       $d$ of the stars which  is  $2.3 -1=1.3$ ,
       see~\cite{Kroupa2001}.
\item  The parameter $a$ varies between 1.3 and 2.7.
\end{enumerate}

The theoretical number of galaxies  as a  function of the red-shift 
presents a maximum that is a function 
of $\alpha$ and $f$   for the 
Schechter  function  
and $c$ , $a$ and $f$ 
for the first ${\mathcal M}-L$  function ;
the agreement with the maximum in the observed 
number of galaxies
is acceptable.

 The observable range in masses can be parameterised as a function
       of $z$  and the ratio between maximum and minimum luminosity
       is  232 at $z$ =0.001 but drops to  1.1  at $z$ =0.15~,
       see  Table~\ref{para_limiting}.

\begin{figure}
\begin{center}
\plotone{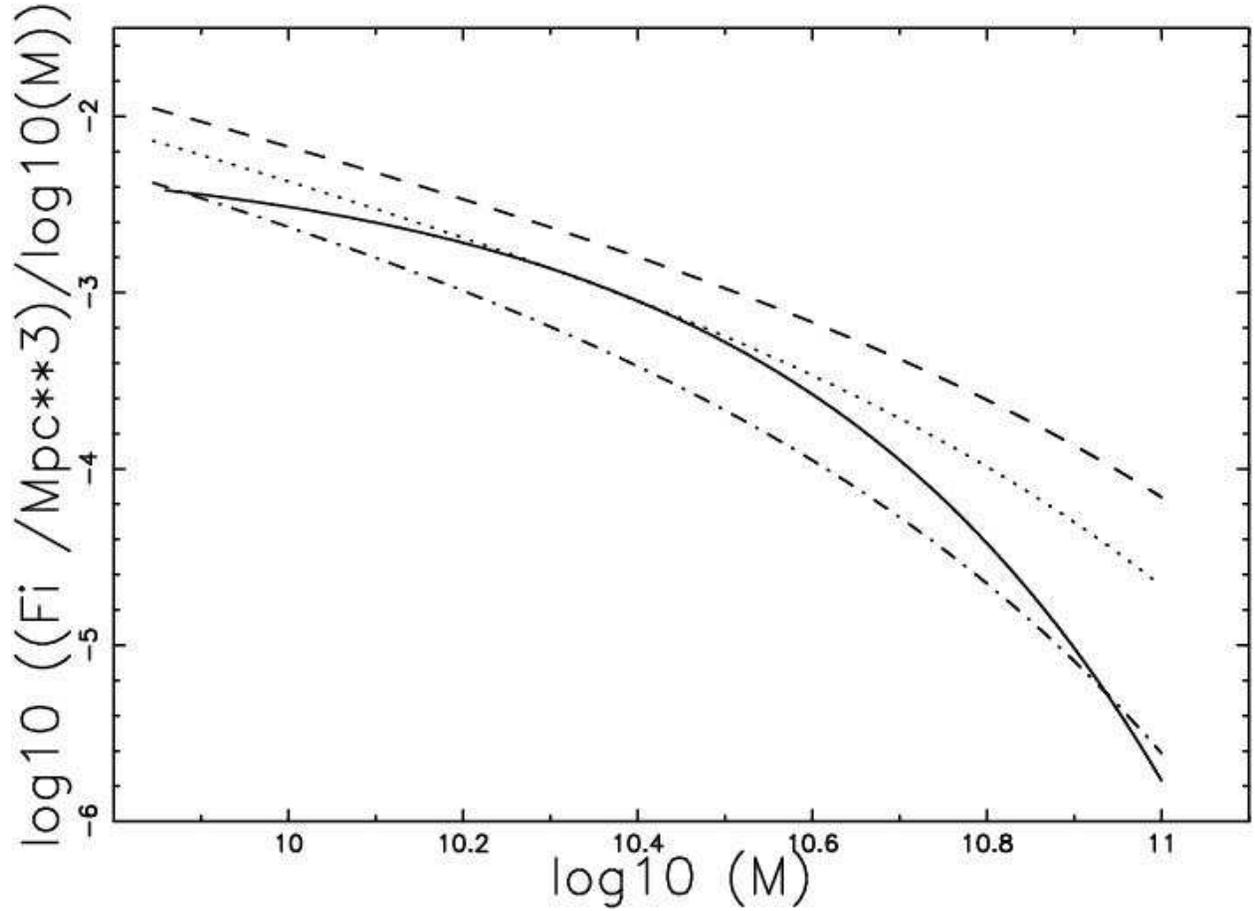} 
\end {center}
\caption {
Barionic Mass function of galaxies:
our  $\Psi ({\mathcal M })/Log_{10}({\mathcal M})$ (full line) , 
                    Bell case            (dashed), 
                    Bottema case         (dot-dash-dot-dash)
                    and Kennicutt-Kroupa case (dotted).
In our case we considered the  SDSS band
  $u^*$ with ${M}_{bol,\sun}$=${M}_{u^*\sun}$=6.39
and $R$=6.0~. 
}
          \label{massedis}%
    \end{figure}

Perhaps a more comprehensive way of comparing the mass estimates
of equation~(\ref{eqnfisher})
( Tully-Fisher relation )
 with those given here (equation~(\ref{massgalaxy})
 and equation~(\ref{massparticular})
) would be as follows.
We take the same sample of galaxies for which the luminosity function was
computed in Section~\ref{test} 
 and compute their mass function  according to a given 
value of $R$ .
This mass function
can be compared with those of other galaxies in a common passband
and, also in this case, the distribution is expressed
through  a  Schechter  function. Three cases are now analysed 
\begin{enumerate}
\item The Bell   case , see \cite{Bell2003b},
where $\Phi^* =0.01 Mpc^{-3}/Log_{10}({\mathcal M})$,
      ${\mathcal M}^*=5.3~10^{10}{\mathcal M_{\sun}}$  
and  $\alpha=-1.21$.
\item The Bottema  case , see \cite{Bottema1997},
where $\Phi^* =0.014 Mpc^{-3}/Log_{10}({\mathcal M})$,
      ${\mathcal M}^*=2.24~10^{10}{\mathcal M_{\sun}}$  
and  $\alpha=-1.20$.
\item Kennicutt-Kroupa case , see \cite{Kennicutt1983,Kroupa1993},
where $\Phi^* =0.011 Mpc^{-3}/Log_{10}({\mathcal M})$,
      ${\mathcal M}^*=3.78~10^{10}{\mathcal M_{\sun}}$  
and  $\alpha=-1.22$. 
\end{enumerate}
Figure~\ref{massedis} reports the already cited standard 
distributions as well as our  
$\Psi ({\mathcal M })/Log_{10}({\mathcal M})$
when the range  in masses  is that given by the conversion
from luminosity to masses.

\clearpage
The analysis of the two new functions for the luminosity for galaxies 
here derived gives a marginally better fit but certainly  
the  Schechter function for it's simplicity and fewer parameters 
still represents a good model for the luminosity function 
for  galaxies.
At present   the study of the Schechter function
is not yet terminated  and two new equations were derived:
equation~(\ref{maximummag}) that represents the maximum 
in magnitude distribution and equation~(\ref{nfunctionz}) 
that gives the value of $z$ at which the observed number 
of galaxies is maximum.
\appendix

\section{On the Kiang Function}
\label{appendixa}

The starting point is the distribution in length , $s$ ,
of a segment in a random fragmentation
\begin{equation}
p(s) = \lambda \exp {(-\lambda s)} ds
\quad ,
\end{equation}
where $\lambda$ is the hazard rate of the exponential  distribution.
Given  the fact that the sum , $u$ , of two  exponential
distributions is
\begin{equation}
p(u)= \lambda^2 u \exp{(- \lambda u)} du
\quad .
\end{equation}
The distribution of 1D Voronoi segments , $l$,
( the midpoint of the sum of two segments) can be found
from the previous formula  by inserting  $u=2l$
\begin{equation}
p(l) = 2 \lambda l \exp{(-2 \lambda l)} d (2 \lambda l)
\quad .
\end{equation}
On transforming in normalised units $x=\frac{l}{\lambda}$
we  obtain
\begin{equation}
p(x) = 2 x \exp{(-2 x) } d (2 x)
\quad .
\end{equation}
After one century  of studies on the Voronoi diagrams
, see the two memories \cite{voronoi_1907} and  \cite{voronoi} , 
the law of the segments in 1D
is the unique analytical result on the field.
When this result is expressed as a gamma variate
we obtain formula(5) of  \cite{kiang}
\begin{equation}
 H (x ;c ) = \frac {c} {\Gamma (c)} (cx )^{c-1} \exp(-cx)
\quad ,
\label{kiang}
\end{equation}
where $  0 \leq x < \infty $ , $ c~>0$
and  $\Gamma (c)$ is the gamma function with argument c;
in the case of 1D Voronoi diagrams $c=2$.
It was conjectured that the area in 2D and the volumes
in 3D of the Voronoi diagrams may be approximated
as the sum of two and three gamma variate of argument 2.
Due to the fact that the sum of n independent gamma
variates with shape parameter $c_i$ is a gamma variate with
shape parameter $c = \sum_{i}^{n} c_i$,
the area and the volumes are supposed to follow a gamma variate
of argument 4 and 6.
This hypothesis was later named
 "Kiang's conjecture", and the equation (\ref{kiang}) used as
a fitting function , see \citet{kumar,Zaninetti2006},
or as  an hypothesis to accept or to reject  using the standard
procedures of the data analysis,
see~\citet{Tanemura1988,Tanemura2003}.
A new way to  parametrise  the 1D, 2D and 3D cells on the
base of the considered dimensionality has been introduced ,
see formula~(12) in \cite{Ferenc_2007}.
\section{On the Truncated  Pareto  Distribution}
\label{appendixb}

The starting  pdf (probability density function)
is the  Pareto distribution~\cite{Pareto,evans}, P,
\begin {equation}
P(x;a,c) = \frac {c a^c}{x^{c+1}} \quad ,
\label{pareto}
\end {equation}
where $  a \leq x < \infty $ , $ a~>0$ , $ c~>0$.
The average value is
\begin{equation}
\overline {x}=
\frac {ca} {c-1}
\quad ,
\end{equation}
which is defined for $c >1$,
and the variance  is
\begin{equation}
\sigma^2 =
{\frac {{a}^{2}c}{ \left(c -2 \right)  \left(c -1 \right) ^{2}}}
\quad ,
\end{equation}
which is defined for $c >2$.
The presence  of an upper bound  ,$b$,   allows us  to
introduce the  following pdf  , named truncated   Pareto $P_T$ ,

\begin {equation}
P_T(x;a,b,c ) = \frac {1}{1-(\frac{a}{b})^c}   \frac {c
a^c}{x^{c+1}} \quad ,  \label{eq:pdf}
\end {equation}

here $  a \leq x \leq b  $ , $ a~>0$ ,$b~>0$, $b>a$ and  $ c~>0$.
The distribution function of the truncated Pareto
is
\begin{equation}
F(x;a,b,c) =
\frac {1 -(\frac {a}{x})^c} {1-(\frac{a}{b})^c}
\quad .
\end{equation}

The  average value of the truncated Pareto pdf is
\begin{equation}
\overline {x} =
\frac {ca} {c-1}
\frac {1 -(\frac{a}{b})^{c-1} } {1-(\frac{a}{b})^c}
\quad ,
\end {equation}
and the variance of the truncated Pareto pdf is
\begin {equation}
 \sigma^2 = \frac {numerator}{denominator}
\quad  ,
\end {equation}
with
\begin{equation}
denominator =
\left( c-2 \right)  \left( c-1 \right) ^{2} \left(
{a}^{2\,c}-2\,{b}^{c}{a}^{c}+{b}^{2\,c} \right)
\quad ,
\nonumber
\end {equation}

\begin{eqnarray}
numerator = \nonumber  \\
c{b}^{2}{a}^{2\,c}+2\,{a}^{2+c}{c}^{2}{b}^{c}-4\,{c}^{2}{a}^{c+1}{b}^{
c+1}+2\,{c}^{2}{b}^{2+c}{a}^{c}-{a}^{2+c}c{b}^{c}
\nonumber \\
+2\,{c}^{3}{a}^{c+1}{
b}^{c+1}-{a}^{2+c}{c}^{3}{b}^{c}-{c}^{3}{b}^{2+c}{a}^{c}-c{b}^{2+c}{a}
^{c}+c{a}^{2}{b}^{2\,c}
\quad .
\nonumber
\end {eqnarray}

The variance of the truncated Pareto
is always defined for every value of $c  >$ 0;
conversely the variance of the
Pareto distribution  can be defined  only when  $c  >$ 2.
The parameter $c$ can be derived through the maximum
likelihood estimator (MLE).
The likelihood function is defined as the probability
 we would have obtained a given set of observations 
if given a
particular set of values of the distribution parameters,$c_i$,
\begin{equation}
L(c) = f (x_1 \ldots x_n  | c_1 \ldots c_n)
\quad .
\end {equation}
If we assume that the n random variables
are independently and
identically distributed,
then we may write the likelihood function as
\begin{equation}
L(c) = f (x_1 |  c_1 \ldots c_p) \ldots
f (x_n| c_1 \ldots c_p) =
\prod_{i=1}^n f (x_i | c_1 \ldots c_p)
\quad .
\end {equation}

The maximum likelihood estimates for the
$c_i$  are
obtained by maximising
the likelihood function, L(c).
Equivalently, we may  find it easier to
maximise  $ln  f(x_i)$ ,
termed the log-likelihood.
So, for a random sample $x_1 \ldots  x_n$
from a truncated Pareto  distribution, the
likelihood function is
given by
\begin{equation}
L(c)  =
\prod_{i=1}^n
c \left(  \left( {a}^{c} \right) ^{-1}- \left( {b}^{c} \right) ^{-1}
 \right) ^{-1} \left( {x_i}^{c+1} \right) ^{-1}
\quad .
\end {equation}
In this model  we have assumed
 that   $a$= min($x_1 \ldots  x_n$) and
$b$= max($x_1 \ldots  x_n$).

Using logarithms, we obtain the log-likelihood
\begin {equation}
\ln L(c) =
nc \ln (a) + n \ln \frac {c}{1 - (\frac{a}{b})^c}
- \sum _{i=1}^n  \ln x_i
\quad .
\end {equation}
Taking the first derivative , we get
\begin {eqnarray}
\frac {\partial }{\partial c}   \ln L(c)  = 0    \nonumber \\
n \ln a       +
\frac {n} {c} +
\frac {n  (\frac {a}{b})^c \ln (\frac{a}{b}) } { 1-(\frac {a}{b})^c  }
- \sum_{i=1}^n  \ln x_i
= 0
\quad .
\end {eqnarray}
The parameter $c$ can be found by solving numerically
the previous  non-linear equation.

\begin{acknowledgements}
I thank Massimo Ramella , Tao Kiang   and the
anonymous referee for  comments.

\end{acknowledgements}

\end{document}